\documentclass[12pt]{iopart}

\usepackage{graphicx}
\usepackage{amssymb}
\usepackage[mathscr]{eucal}

\newcommand{\email}[1]{\ead{#1}}
\newcommand{\affiliation}[1]{\address{#1}}
\newcommand{\acknowledgments}{\ack}

\newcommand{\sss}[1]{{\scriptscriptstyle{#1}}}

\newcommand{\uPl}{\mathrm{Pl}}

\newcommand{\uS}{\mathrm{S}}

\newcommand{\usssS}{\sss{\uS}}

\newcommand{\usssPl}{\sss{\uPl}}

\newcommand{\nS}{n_\usssS}

\newcommand{\ie}{\textrm{i.e.}~}

\newcommand{\mpl}{m_\usssPl}

\newcommand{\bear}{\begin{array}}  \newcommand{\eear}{\end{array}}
\newcommand{\bea}{\begin{eqnarray}}  \newcommand{\eea}{\end{eqnarray}}
\newcommand{\beq}{\begin{equation}}  \newcommand{\eeq}{\end{equation}}
\newcommand{\bef}{\begin{figure}}  \newcommand{\eef}{\end{figure}}
\newcommand{\bec}{\begin{center}}  \newcommand{\eec}{\end{center}}

\newcommand{\lkk}{\left[}  \newcommand{\rkk}{\right]}

\def\Vec#1{{\mbox{\boldmath$#1$\unboldmath}}}
\def\sVec#1{{\mbox{\boldmath$\scriptstyle{#1}$\unboldmath}}}

\def\setN{\mathbb{N}}

\begin{document}

\title{Generation of Large-Scale Magnetic Fields in Single-Field
Inflation}

\author{J\'er\^ome Martin} \email{jmartin@iap.fr} \affiliation{ Institut
d'Astrophysique de Paris, UMR 7095-CNRS, Universit\'e Pierre et Marie
Curie, 98bis boulevard Arago, 75014 Paris, France}

\author{Jun'ichi Yokoyama} \email{yokoyama@resceu.s.u-tokyo.ac.jp}
\affiliation{Research Center for the Early Universe, Graduate School of
Science, The University of Tokyo, Tokyo 113-0033, Japan}

\date{\today}

\begin{abstract}
We consider the generation of large-scale magnetic fields in slow-roll
inflation. The inflaton field is described in a supergravity framework
where the conformal invariance of the electromagnetic field is
generically and naturally broken. For each class of inflationary
scenarios, we determine the functional dependence of the gauge coupling
that is consistent with the observations on the magnetic field strength
at various astrophysical scales and, at the same time, avoid a
back-reaction problem. Then, we study whether the required coupling
functions can naturally emerge in well-motivated, possibly
string-inspired, models. We argue that this is non trivial and can be
realized only for a restricted class of scenarios. This includes
power-law inflation where the inflaton field is interpreted as a
modulus. However, this scenario seems to be consistent only if the
energy scale of inflation is low and the reheating stage
prolonged. Another reasonable possibility appears to be small field
models since no back-reaction problem is present in this case but,
unfortunately, the corresponding scenario cannot be justified in a
stringy framework. Finally, large field models do not lead to sensible
model building.
\end{abstract}

 \maketitle

\section{Introduction}
\label{sec:introduction}

Magnetic fields are present on various scales in the Universe from
planets, stars, galaxies, to clusters of galaxies, see {\it e.g.},
Refs.~\cite{Kronberg:1993vk, Grasso:2000wj, Widrow:2002ud} for
reviews. In galaxies of all types, magnetic fields with the field
strength $\sim 10^{-6}$ G, ordered on $1-10$ kpc scale, have been
detected \cite{Widrow:2002ud,Sofue:1986wc}. There is also some evidence
that they exist in galaxies at cosmological
distances~\cite{Kronberg:1992}.  Furthermore, in recent years, magnetic
fields in clusters of galaxies have been observed by means of the
Faraday rotation measurements (RMs) of polarized electromagnetic
radiation passing through an ionized
medium~\cite{Kim:1990,Kim:1991,Clarke:2001}. Unfortunately, however, RMs
give only the product of the field strength along the line of sight and
the coherence scale, and we cannot find them independently.  In this
situation, the strength and the scale are estimated on $10^{-7}-10^{-6}$
G and 10 kpc$-$1 Mpc, respectively.  It is interesting that magnetic
fields in clusters of galaxies are as strong as galactic ones and that
the coherence scale may be as large as $\sim$ Mpc.

\par

Since the conductivity of the Universe through most of its history is
large, the magnetic field $B$ evolves conserving magnetic flux as $B
\propto a^{-2}$, where $a(t)$ is the scale factor. On the other hand,
the average cosmic energy density $\bar{\rho}$ evolves as $\bar{\rho}
\propto a^{-3}$. Hence, the scaling of the magnetic field is given by
$B \propto {\bar{\rho}}^{2/3}$ and this relation should also be
understood as giving the value of $B$ at different spatial locations
characterized by different energy densities. The present ratio of
interstellar medium density in galaxies ${\rho}_\mathrm{gal}$ to
$\bar{\rho}$ and that of inter-cluster medium density in clusters of
galaxies ${\rho}_\mathrm{cl}$ are ${\rho}_\mathrm{gal}/\bar{\rho}
\simeq 10^5-10^6$ and ${\rho}_\mathrm{cl}/\bar{\rho} \simeq
10^2-10^3$, respectively.  Consequently, from these relations, we see
that, using the law mentioned before, namely $\bar{B}=B_{\mathrm{gal,
cl}} (\rho _{\mathrm{gal,cl}}/\bar{\rho})^{-2/3}$, the required
strength of the cosmic magnetic field at the structure formation,
adiabatically rescaled to the present time, is $10^{-10}-10^{-9}
\mathrm{G}$ in order to explain the observed fields mentioned
before. On the other hand, in general, if the galactic dynamo
mechanism is at play, then seed fields with a present strength of only
$10^{-22}-10^{-16} \mathrm{G}$ is
required~\cite{ParkerEN:1970,ParkerEN:1979,Zelodovitch:1983}.

\par

The question addressed in this article is that of the origin of these
magnetic fields and whether there exists simple, physically
well-motivated mechanisms, which could provide the target values of
$B$ just described above. Generation mechanisms of the seed magnetic
fields fall into two broad categories. One is astrophysical processes
and the other is cosmological physical processes in the early
Universe. The former exploits the difference in mobility between
electrons and ions. This difference can lead to electric currents and
hence magnetic fields. If the scale of cluster magnetic fields is as
large as $\sim$ Mpc, however, it is unlikely that the origin of such a
large-scale magnetic field is in astrophysical processes and we are
tempted to assign the origin of cosmological magnetic field to the
processes during inflation in the early Universe which stretches
coherent scales
exponentially~\cite{Guth:1980zm,Sato:1980yn,Linde:2005ht,Olive:1989nu,
Kolb:1990aa,Lyth:1998xn, Martin:2003bt,Martin:2004um,Martin:2007bw}.

\par

During inflation quantum scalar (density)
perturbations~\cite{Mukhanov:1981xt,Hawking:1982cz, Starobinsky:1982ee,
Guth:1982ec} and tensor (gravitational waves)
perturbations~\cite{Grishchuk:1974ny,Rubakov:1982df} are generated and
stretched to cosmological scales by subsequent accelerated expansion.
Usually, however, large scale fluctuations in electromagnetic fields are
not generated, because they are conformally
invariant~\cite{Parker:1968}. Hence, in order to generate cosmological
magnetic fields during inflation, one must break the conformal
invariance of the theory in an appropriate manner.  A number of models
have been proposed in this context so
far~\cite{Turner:1987bw,Ratra:1991av,Ratra:1991bn,Lemoine:1995vj,
Giovannini:2001nh,Giovannini:2001xk,Giovannini:2002ki,Calzetta:1997ku,
Kandus:1999st,Davis:2000zp,Dolgov:1993vg,Bamba:2003av,Bamba:2004cu}.

\par

Here we propose a simple model making use of the arbitrariness of the
gauge kinetic function in supergravity~\cite{Bailin:1994}. Indeed, the
general supergravity Lagrangian including vector super-multiplets
contains the following term
\begin{equation}
\int {\rm d}^2\theta \left[f_{ab}(\Phi)W_a{}^{\alpha }W_{\alpha b}
+\mbox{h.c.}\right]\, ,
\end{equation}
where $W_a{}^{\alpha }$ is the gauge field strength superfield with
spinor index $\alpha $ and gauge group index $a$. As already mentioned,
the quantity $f_{ab}(\Phi)$ is an arbitrary function of the chiral
superfield $\Phi$~\cite{Cremmer:1982en}. This implies that the conformal
invariance is naturally and generically broken as is most conveniently
shown when the above term is expressed in terms of the superfields
components
\begin{equation}
\frac14 \Re f_{ab}(\phi)\left(F_{a}\right)_{\mu \nu}
\left(F_{b}\right)^{\mu \nu}\, .
\end{equation}
Since we are interested in the production of magnetic fields, we
consider a $U(1)$ gauge field and, therefore, the gauge field strength
do no longer carry a gauge index. As a consequence, there is also only
one function $f(\phi)$ left. Moreover, in order to consider the
minimalist case with the least possible assumptions, we naturally
identify the scalar part $\phi $ of the chiral superfield $\Phi$ to
the inflaton field.

\par

A related study was carried out in
Refs.~\cite{Ratra:1991av,Ratra:1991bn} but only in the particular case
where $f(\phi )\propto {\rm e}^{\phi}$. Here, we study the most general
case (at least, in the framework of single field inflation) and
determine which form $f(\phi )$ should take in order to obtain a power
spectrum of the magnetic field in agreement with the currently available
observational constraints on large scales. Then, for each category of
single field inflationary scenarios, we explore whether there is a
natural model in supergravity (possibly string-inspired) that could
reproduce the required shape of the gauge coupling and, at the same
time, leads to the correct form of the inflaton potential.

\par

This paper is organized as follows. In Sec.~\ref{sec:reverse}, we
describe our model action and derive the equations of motion from it.
Then, the gauge field is quantized in the Coulomb gauge and the vacuum
energy density of the magnetic field at the end of inflation is
determined. The evolution of the magnetic energy density through the
subsequent stages of evolution, preheating, radiation and matter
dominated eras is also computed. In a next step, for each single field
slow-roll inflationary models, using a simple ansatz depending only on
one free parameter, we calculate the required shape of the gauge
coupling function $f(\phi )$. In Sec.~\ref{sec:couplingconstraint}, we
use the currently available limits on the magnetic field to constraint
the free parameter mentioned above, that is to say to constraint the
shape of the gauge coupling function. We also investigate the
consistency of the model and determine under which conditions this one
does not suffer from a back-reaction problem. In
Sec.~\ref{sec:particlemodels}, we try to embed in particle physics the
models that have been shown to be phenomenologically viable. Finally, in
Sec.~\ref{sec:conclusion}, we present our conclusions. Throughout this
article, we use units in which $ c =\hbar = M_G=k_\mathrm{B} =1$. We
also put the reduced Planck mass, $M_G\equiv 1/\sqrt{8\pi G}$, to unity
when it simplifies the notation. In this case, the full Planck scale
takes $\mpl \equiv G^{-1/2}=\sqrt{8\pi}$.

\section{Reverse-Engineering of the Magnetic Field}
\label{sec:reverse}

\subsection{Magnetic Field during Inflation}
\label{sec:magneticduringinflation}

During inflation, following the considerations presented in
Sec.~\ref{sec:introduction}, the action of the system, namely a scalar
inflaton field plus a $U(1)$ gauge field, can be written as
\begin{eqnarray}
\label{eq:action} 
S\left[\phi, A_{\mu }\right] &=& -\int {\rm d}^4x
\sqrt{-g}\left[ \frac12g^{\mu \nu} \partial _{\mu}\phi \partial
_{\nu}\phi+V(\phi)\right] \nonumber \\ & & 
-\frac14\int {\rm d}^4x \sqrt{-g}g^{\alpha
\beta }g^{\mu \nu} f^2\left(\phi \right)F_{\mu \alpha }F_{\nu \beta }\,
, \label{action}
\end{eqnarray}
with $F_{\mu \alpha}\equiv \partial _{\mu }A_{\alpha}-\partial
_{\alpha }A_{\mu}$, $A_{\alpha}$ being the vector potential whose
dimension is that of a mass $[A_{\mu}]=M$. The inflaton potential
$V(\phi)$ and the dimensionless gauge coupling $f(\phi)$ are, a
priori, arbitrary functions. Notice that we have changed the
definition of the gauge coupling (now $f^2$ appears in the action
rather than $f$) for future convenience. The equations of motion
following from the above action can be written as
\begin{eqnarray}
\frac{1}{\sqrt{-g}}{\partial}_{\mu} \left(
\sqrt{-g}g^{\mu\nu}{\partial}_{\nu} \phi \right) -\frac{{\rm d}V}{{\rm
d}\phi} &=& \frac{1}{4}\frac{{\rm d}f(\phi)}{{\rm d}\phi}
F_{\mu\nu}F^{\mu\nu}\, ,\\ 
\partial _{\mu}\left[\sqrt{-g}g^{\mu \nu}g^{\alpha \beta
}f^2\left(\phi \right)F_{\nu \beta }\right]&=&0\, .
\end{eqnarray} 
In the following, we consider that the electromagnetic field is a
perturbatively small quantity (a ``test'' field) compared with the
inflaton counterpart. This means that one can neglect the right hand
side of the Klein-Gordon equation and work in a spatially flat
Friedmann-Lema\^{\i}tre-Robertson-Walker (FLRW) space-time,
\begin{eqnarray}
\label{eq:metric}
{\rm d}s^2 = g_{\mu\nu}{\rm d}x^{\mu}{\rm d}x^{\nu} = -{\rm d}t^2 +
a^2(t){\rm d}{\Vec{x}}^2= a^2(\eta)\left(-{\rm d}\eta^2 + {\rm
d}{\Vec{x}}^2\right)\, , 
\end{eqnarray} 
where $t$ is the cosmic time and $\eta $ the conformal one, since the
gauge field being negligible there is no preferred
direction. Therefore, the equations of motion are just the standard
ones, namely
\begin{eqnarray}
\label{eq:friedman}
H^2 &=& \left( \frac{\dot{a}}{a} \right)^2 = \frac{1}{3}
\left[\frac{1}{2}{\dot{\phi}}^2 + V(\phi)\right] \, ,\quad
\frac{1}{a^2}{\cal H}^2 = \frac{1}{3}\left[\frac{1}{2a^2}{\phi '}^2 +
V(\phi)\right]\, , 
\end{eqnarray}
with ${\cal H}\equiv a'/a$ and, for the inflaton field, 
\begin{eqnarray}
\label{eq:eqmotioncosmicphi} 
\ddot{\phi} + 3H\dot{\phi} + \frac{{\rm d}V}{{\rm
d}\phi} = 0\, ,\quad \phi '' + 2{\cal H}\phi '+ \frac{{\rm d}V}{{\rm
d}\phi} = 0\, .
\end{eqnarray}
In order to simplify the calculations, we adopt the Coulomb gauge
where $A_0(t,\Vec{x}) = 0$ and ${\partial}_jA^j(t,\Vec{x}) =0$. In
this case, one obtains
\begin{eqnarray}
\label{eq:eqmotioncosmicA} 
\ddot{A_i}(t,\Vec{x}) + \left( H +
2\frac{\dot{f}}{f} \right) \dot{A_i}(t,\Vec{x}) -
{\partial}_j{\partial}^jA_i(t,\Vec{x}) &=& 0\, ,\\
A_i ''+2\frac{f'}{f}A_i'-a^2\partial _j\partial
^jA_i &=& 0\, ,
\end{eqnarray} 
where, of course, $\partial ^j=g^{jk}\partial _k=(\delta
^{jk}/a^2)\partial _k$. If one defines $\bar{A}_j\equiv f A_j$, then
one can absorb the part proportional to the first derivative of the
vector potential and put the equation of motion under the standard
form of an equation for an oscillator. One gets
\begin{equation}
\bar{A}_i''-\frac{f''}{f}\bar{A}_i-a^2\partial _j\partial
^j\bar{A}_i=0\, .
\end{equation}
We see that, if $f=1$, the equation of motion of $A^j$ reduces to that
of an harmonic oscillator. Due to the conformal coupling of the gauge
sector, the dynamics is non trivial only if $f$ is not a constant.

\par

Let us now (covariantly) define the electric and magnetic fields seen
by an observer characterized by the four-velocity vector $u^{\mu
}$. One has~\cite{Barrow:2006ch}
\begin{equation}
E_{\mu }=u^{\nu }F_{\mu \nu}\, ,\quad B_{\mu }=\frac12 \epsilon _{\mu \nu
 \kappa}F^{\nu \kappa}\, ,
\end{equation}
where the tensor $\epsilon _{\mu \nu \kappa}$ is defined by the
relation
\begin{equation}
\epsilon _{\mu \nu \kappa}=\eta _{\mu \nu \kappa \lambda }u^{\lambda }\,
 ,
\end{equation}
$\eta _{\mu \nu \kappa \lambda }$ being the totally antisymmetric
permutation tensor of space-time with $\eta ^{0123}=(-g)^{-1/2}$ or
$\eta _{0123}=(-g)^{1/2}$. Therefore, for a comoving observer with
$u^{\mu }=\left(1, \Vec{0}\right)$ (in cosmic time), one gets
\begin{equation}
E_i=-\dot{A}_i\, ,\quad B_i
=\frac{1}{a}\epsilon _{ijk}\partial _jA_k \, ,
\end{equation}
with $\epsilon _{123}=1$. A remark is in order concerning the formula
giving the magnetic field. Contrary to the previous expressions, it is
not written in a covariant way and, therefore, the position of the
indices is irrelevant even if implicit summation is still present. So,
for instance, this formula simply means $B_1=(\partial _2A_3-\partial
_3A_2)/a$. As expected, the dimension of the magnetic field is two,
$[B_i]=M^2$.

\par

We now turn to the quantization of the system. It follows from
Eq.~(\ref{action}) that the momentum conjugate to the gauge field
$A_{i}(t,\Vec{x})$ is given by
\begin{eqnarray}
\label{eq:defpi}
\pi ^{i}(t,\Vec{x}) =\frac{\delta S}{\delta \dot{A}_i}
= f^2(\phi)a^3(t) g^{ij}\dot{A_j}(t,\Vec{x})
= f^2(\phi)a^2(\eta )g^{ij}A_j'(\eta ,\Vec{x})\, ,
\end{eqnarray}
which also implies that $\pi _{i}(t,\Vec{x})=f^2(\phi)a^3(t)
\dot{A_i}(t,\Vec{x})$. For quantization, we impose the canonical
commutation relation between $A^i(t,\Vec{x})$ and
${\pi}_{j}(t,\Vec{x})$:
\begin{eqnarray}
\label{eq:commutperp} 
\left[A^i(t,\Vec{x}), {\pi}_{j}(t,\Vec{y})
\right] &=& i \int \frac{{\rm d}^3 \Vec{k}}{{(2\pi)}^{3}}
{\rm e}^{i \sVec{k} \cdot \left( \sVec{x} - \sVec{y} \right)} \left(
{\delta}^{i}{}_{j} - \delta _{j\ell}\frac{k^{i} k^{\ell}}{k^2 } \right)\\
&=&
i\delta ^{(3)}_{\perp }{}^i{}_j\left(
\Vec{x}-\Vec{y}\right)\, , 
\end{eqnarray}
where $\delta ^{(3)}_{\perp ij}$ is the transverse delta function
introduced in order to have a consistent quantization in the Coulomb
gauge.

\par

In order to define the creation and annihilation operators, one has to
Fourier expand the vector potential operator. For this purpose, one
introduces an orthonormal basis in space-time defined by (in conformal
time)
\begin{equation}
\label{eq:defepsilon}
\epsilon _0^{\mu }=\left(\frac{1}{a}, \Vec{0}\right)\, ,\quad 
\epsilon _\lambda ^{\mu}=\left(0,
\frac{\tilde{\epsilon }_{\lambda }^i}{a}\right)
\, ,\quad \epsilon _3^{\mu}=\left(0,\frac{1}{a}\frac{k^i}{k}\right)\, ,
\end{equation}
where $\lambda =1,2$ and where, by definition, $\delta
_{ij}\tilde{\epsilon }_{\lambda }^i \tilde{\epsilon }_{\lambda
}^j\equiv 1 $ (no summation on $\lambda$). In the above equations,
$k^i=a k^i_{\rm phys}$ is the comoving wavenumber satisfying $\delta
_{ij}k^ik^j=k^2$. In addition, $\epsilon _{\lambda
}^{\mu}\epsilon_{3\mu }= \tilde{\epsilon}_{\lambda }^ik_i=0$ which
expresses the Coulomb gauge in Fourier space. Notice that, without the
factor $1/a$ in the previous definitions, the above vectors would not
be correctly normalized in curved space-time. Then, one has the
covariant completeness relation $\sum _{s=0}^{s=3}g_{ss}\epsilon
^{\mu}_{s}\epsilon ^{\nu}_s =g^{\mu \nu}$ where $g_{ss}$ is just a
factor which stands for $+1$ or $-1$. Projected on space, the
completeness relation reduces to
\begin{equation}
\label{eq:completspace}
\sum _{\lambda =1}^2\epsilon ^{i}_{\lambda } (\Vec{k})
\epsilon _{j \lambda }(\Vec{k})
+\delta _{j\ell }\frac{k^ik^{\ell}}{k^2}=\delta ^i{}_j\, .
\end{equation}
Finally, the expression of $A_i(t,\Vec{x})$ can be written as
\begin{eqnarray} 
\label{eq:FourierA}
  A_i(\eta ,\Vec{x}) &=&
\int \frac{{\rm d}^3 k}{(2\pi)^{3/2}}\sum _{\lambda =1}^2
\epsilon _{i \lambda }(\Vec{k})\biggl[
b_{\lambda} (\Vec{k}) A(\eta,k){\rm e}^{i \sVec{k} \cdot
  \sVec{x} } \nonumber \\ & & 
+ b_{\lambda }^{\dagger}(\Vec{k}) {A}^*(\eta ,k){\rm e}^{-i
  \sVec{k} \cdot \sVec{x}} \biggr], 
\end{eqnarray} 
in terms of the annihilation and creation operators,
$b_{\lambda}(\Vec{k})$ and $b_{\lambda}^{\dagger}(\Vec{k})$, with
$\Vec{k}$ being the comoving wavenumber. The quantity $\epsilon
_i^\lambda (\Vec{k})$ is the transverse polarization vector and has
been defined before and the time-dependent Fourier amplitude obeys the
following equation of motion
\begin{equation}
\label{eq:fourieramplitude}
{\cal A}''(\eta ,k)+\left(k^2-\frac{f''}{f}\right){\cal A}(\eta ,k)=0
\end{equation}
where ${\cal A}(\eta ,k)\equiv a(\eta )f(\eta )A(\eta ,k) $. The
additional factor $a$ in the previous definition originates from the
presence of the vector $\epsilon _{i\lambda }(\Vec{k})$ in the Fourier
expansion~(\ref{eq:FourierA}) of the gauge field. The dimension of the
Fourier amplitude of the gauge field is $[A]=M^{-1/2}$ since $[A_i]=M$
and $[b_{\lambda}(\Vec{k})]=M^{3/2}$ (because the delta function, see
below, has dimension equal to the inverse of its argument). In order
to satisfy Eq.~(\ref{eq:commutperp}), the creation and annihilation
operators must satisfy
\begin{eqnarray}
\label{eq:comutb} 
\left[b_{\lambda}(\Vec{k}), b_{\lambda '}^{\dagger}({\Vec{k}}^{\prime})
\right] = {\delta}^3
(\Vec{k}-{\Vec{k}}^{\prime})\delta_{\lambda \lambda '}\, , \quad \left[ 
b_{\lambda }(\Vec{k}), b_{\lambda '}({\Vec{k}}^{\prime})\right]
= \left[b_{\lambda }^{\dagger}(\Vec{k}),
b_{\lambda '}^{\dagger}({\Vec{k}}^{\prime})\right] = 0\, ,\nonumber \\
\end{eqnarray}
where we have used Eq.~(\ref{eq:completspace}). The normalization
condition for the time-dependent amplitude $A(t,k)$ should be chosen
such that
\begin{eqnarray} 
\label{eq:wronskian}
A(t,k){\dot{A}}^{*}(t,k) - {\dot{A}}(t,k){A}^{*}(t,k)
= \frac{i}{f^2a^3}\, .
\end{eqnarray}

Let us now calculate the energy density of the magnetic field. The
stress energy tensor is given by
\begin{equation}
\label{eq:stresstensor}
T_{\mu \nu}\equiv -\frac{2}{\sqrt{-g}}\frac{\delta S}{\delta g^{\mu \nu}}
=-f^2(\phi)g^{\alpha \beta }F_{\mu \alpha }F_{\beta \nu}
-\frac{f^2(\phi)}{4}g_{\mu \nu}g^{\alpha \beta }g^{\gamma \delta }F_{\beta
\delta }F_{\alpha \gamma}\, ,
\end{equation}
and the energy density associated to the ``magnetic'' part can be
expressed as (of course, there are also contributions depending on
$\partial _0A_i$, that is to say depending on the electric field, that
we do not consider here)
\begin{equation}
T_{00}^{B}=\frac{f^2(\phi)}{4}a^2g^{ij}g^{k\ell }
\left(\partial _jA_{\ell}-\partial _{\ell}A_j\right)
\left(\partial _iA_{k}-\partial
 _{k}A_i\right)=\frac{f^2(\phi)}{2}a^2B_{\mu }B^{\mu}\, .
\end{equation}
As expected, one has $[T_{00}^B]=M^4$ thanks to $[A_i]=M$. Using the
Fourier expansion of the potential vector, straightforward
manipulations lead to
\begin{eqnarray}
\left \langle T^{B \, 0}{}_0\right \rangle &=&-
\frac{f^2(\phi)}{4}\frac{1}{(2\pi)^3}\int {\rm d}^3k
\left \vert A(\eta ,k)\right\vert ^2g^{ij}g^{m\ell}
\nonumber \\ & & \times
\sum _{\lambda =1}^2\left(\epsilon _{\ell \lambda }k_j
-\epsilon _{j \lambda }k_{\ell}\right)
\left(\epsilon _{m\lambda }k_i
-\epsilon _{i\lambda }k_{m}\right)\, .
\end{eqnarray}
The final step is to notice that
\begin{eqnarray}
& & g^{ij}g^{m\ell}
\sum _{\lambda =1}^2\left(\epsilon _{\ell \lambda }k_j
-\epsilon _{j \lambda }k_{\ell}\right)
\left(\epsilon _{m\lambda }k_i
-\epsilon _{i\lambda }k_{m}\right)\nonumber \\
&=& 2\frac{k^2}{a^2}\sum _{\lambda =1}^2\epsilon_{\ell \lambda }
\epsilon ^{\ell}_{\lambda}
-\frac{2}{a^2}k_jk^{\ell}\sum _{\lambda =1}^2\epsilon_{\ell \lambda} 
\epsilon ^{j}_{\lambda}=4\frac{k^2}{a^2}\, ,
\end{eqnarray}
where we have used the space component of the completeness
relation. The energy density, $\rho _{_{B}}\equiv -\left \langle
T^{B\, 0}{}_0 \right \rangle$, is therefore given by
\begin{eqnarray}
\rho _{_{B}}(\eta)&=&\frac{1}{2\pi ^2}\int_{0}^{+\infty}\frac{{\rm d}k}{k}k^5
f^2(\phi )\left \vert \frac{A(\eta ,k)}{a(\eta)}\right\vert ^2
\nonumber \\
&=&\frac{1}{2\pi ^2}\int_{0}^{+\infty}\frac{{\rm d}k}{k}k^5
\frac{1}{a^4(\eta)}\left \vert {\cal A}(\eta ,k)\right\vert ^2
\equiv \int_{0}^{+\infty}
{\rm d}k\frac{{\rm d}}{{\rm d}k}\rho_{_{B}}(\eta ,k)\, ,
\end{eqnarray}
or, considering only the energy density stored at a given scale,
\begin{equation}
\label{eq:rhoB}
\frac{{\rm d}}{{\rm d}k}\rho_{_{B}}(\eta ,k)=
\frac{1}{2\pi ^2}\frac{k^4}{a^4(\eta)}
\left \vert {\cal A}(\eta ,k)\right\vert ^2\, .
\end{equation}
As expected, the fourth power of the physical wave number, $k_{\rm
  phys}\equiv k/a(\eta )$ appears in the above expression.

\par

Let us now compute the power spectrum in the case where the scale
factor is given by a power law of the conformal time,
\begin{equation}
a(\eta )=a_0\left\vert \frac{\eta }{\eta _0}\right\vert ^{1+\beta}\, .
\end{equation}
The case $\beta =-2$ corresponds to de Sitter space-time. One can also
show~\cite{Martin:1999wa} that the spectral index of density
perturbations is given by $\nS=2\beta +5$. If we adopt the very
conservative bound $\vert \nS-1\vert \lesssim 0.1$ then this amounts
to choosing the index $\beta $ such that $-2.05\lesssim \beta
<-2$. Notice that the above assumption for the scale factor is not as
restrictive as it may seem at first sight. Indeed, the above law is
also valid in the slow-roll approximation where one
has~\cite{Martin:1999wa}
\begin{equation}
a(\eta )\propto \vert \eta \vert ^{-1-\epsilon
_1}\, , 
\end{equation}
where $\epsilon _1$ is the first slow-roll parameter and is constant
at first order. Therefore, our ansatz allows us to treat the case of
slow-roll inflation as, for instance, large and small field inflation
or even hybrid inflation in the inflationary valley.

\par

Let us now discuss the form of the gauge coupling function $f(\phi)$
that will be considered in this article. One assumes that
\begin{equation}
\label{eq:ansatzf}
f(\eta) \propto a^{\alpha} \, ,
\end{equation}
where $\alpha $ is a free index. The motivation for such a choice is as
follows. Unfortunately, the precise form of the magnetic spectrum is not
known experimentally. Only upper limits on the amplitude at given scales
have been obtained so far, see below. Therefore, one cannot take the
function $\rho _{_{B}}$ and reverse-engineer it exactly in order to find
the corresponding $f(\phi )$. On the other hand, it will be demonstrated
below that the coupling function with $\alpha =2$ leads to a
scale-invariant power spectrum, see also
Ref.~\cite{Bamba:2003av}. Clearly, since the amplitude of the primordial
magnetic field is not strongly peaked over a specific range of scales
(either small or large scales), a scale-invariant spectrum satisfies the
currently available experimental data. Thus, the coupling function
$f\propto a^2$ certainly belongs to a class of models which lead to
interesting scenarios. As a consequence, it seems natural to consider
generalizations of the simple choice $f\propto a^2$. In fact, it will be
proven in the following that the ansatz~(\ref{eq:ansatzf}) leads to a
power-law for the spectrum of the magnetic field, the tilt $n_{_{B}}$
being determined by the value of the parameter $\alpha $ (and
$n_{_{B}}=0$ for $\alpha =2$) and, as a consequence, allows us to treat
a simple but quite generic and general class of models. As mentioned
before, the data are not yet very accurate and are still compatible with
important variations of $\alpha $ around the preferred value $\alpha
=2$. Given the present-day measurements, the corresponding range will be
determined below. Then, given the ansatz~(\ref{eq:ansatzf}), we will
reverse-engineer the spectrum and find the corresponding function
$f(\phi )$ for a given model of inflation, taking into account the
uncertainties on $\alpha $. This will allow us to produce a complete
class of models that satisfies the currently available experimental
data. Then, in a second step, we will seek for particle physics
realizations of these scenarios.

\par

\begin{figure}
\begin{center}
\includegraphics[width=12.5cm]{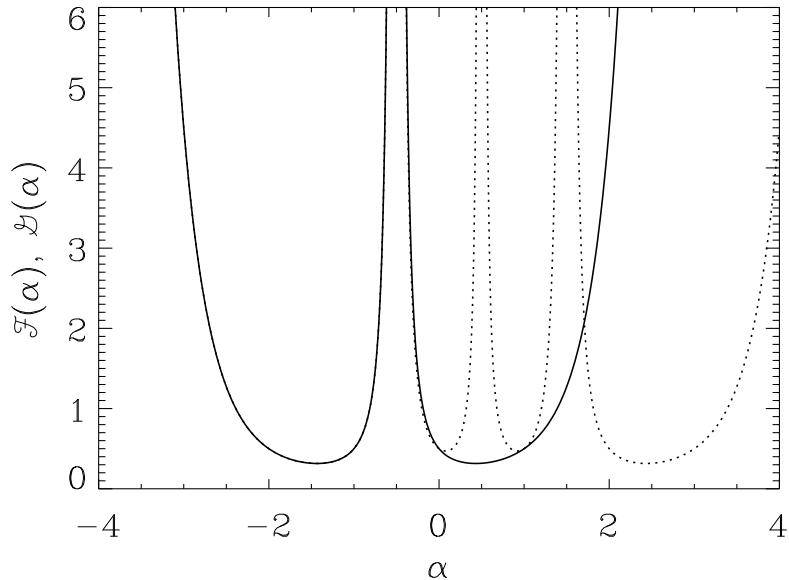} 
\caption{Amplitude of the magnetic power spectrum (solid line) at the
end of inflation, as given by the function ${\cal F}(\alpha )$ defined
in Eq.~(\ref{eq:defF}), in terms of the index $\alpha $ characterizing
the shape of the gauge coupling. An almost scale-invariant power
spectrum, $\beta \simeq -2$, has been assumed such that $\gamma \equiv
\alpha (1+\beta )\simeq -\alpha $. The divergence at $\alpha =-1/2$
signals the transition between the two branches of the spectrum. The
amplitude of the electric power spectrum at the end of inflation is
also displayed (dotted line). As shown in the following, the amplitude
is given by the function ${\cal G}(\alpha )$ defined in
Eq.~(\ref{eq:defG}).}
\label{fig:amplitude}
\end{center}
\end{figure}

Given Eq.~(\ref{eq:ansatzf}), the effective potential $f''/f$ can be
expressed as
\begin{equation}
\frac{f''}{f}=\frac{\gamma(\gamma-1)}{\eta ^2}\, ,
\end{equation}
where we have defined $\gamma \equiv \alpha (1+\beta)$. It is
straightforward to integrate Eq.~(\ref{eq:fourieramplitude}) in terms
of Bessel functions. The result reads
\begin{equation}
\label{eq:sol}
{\cal A}(\eta, k)=(k\eta )^{1/2}\left[C_1(k)J_{\gamma-1/2}(k\eta )
+C_2(k)J_{-\gamma+1/2}(k\eta )\right]\, ,
\end{equation}
where $C_1(k)$ and $C_2(k)$ are two scale-dependent coefficients which
are fixed by the initial conditions. These ones are determined in the
ultra-violet limit. Indeed, in the short-wavelength limit $k/(a H) =
-k\eta \to \infty $ , the vacuum reduces to the one in Minkowski
space-time
\begin{eqnarray}
\label{eq:ciA}
{\cal A} (k,\eta) \to \frac{1}{\sqrt{2k}} {\rm e}^{-ik\eta}\, . 
\end{eqnarray} 
Notice that one can easily check that the normalization is
correct. Indeed, the above relation means $A(\eta ,k)\rightarrow {\rm
e}^{-ik\eta }/(af\sqrt{2k})$ and taking into account the fact that
$\dot{A}=A'/a$, the previous mode function correctly reproduces the
Wronskian given by Eq.~(\ref{eq:wronskian}) (in addition, $[A] =[{\cal
A}]=M^{-1/2}$ as announced previously). As a consequence, the two
arbitrary coefficients $C_1$ and $C_2$ are given by
\begin{equation}
C_1(k)=-C_2(k){\rm e}^{i\pi(\gamma-1/2)}\, ,\quad 
C_2(k)=\sqrt{\frac{\pi}{4k}}\frac{{\rm e}^{-i\pi(\gamma +1)/2}}
{\cos (\pi\gamma )}\, .
\end{equation}
On large scales (compared to the Hubble radius), $k/(a H) = -k\eta \to
0$, using the asymptotic behavior of the Bessel functions, one obtains
\begin{eqnarray}
{\cal A} (k,\eta) &\to& \frac{\sqrt{\pi}}{2^{\gamma+1/2}}
\frac{{\rm e}^{i\pi \gamma /2}}{\Gamma (\gamma+1/2)
\cos (\pi\gamma )}k^{-1/2}(k\eta )^{\gamma}
\nonumber \\
& & +\frac{\sqrt{\pi}}{2^{-\gamma +3/2}}
\frac{{\rm e}^{i\pi (1-\gamma)/2}}{\Gamma (-\gamma+3/2)
\cos \left[\pi(1-\gamma)\right]}k^{-1/2}(k\eta )^{1-\gamma}\, ,
\end{eqnarray}
where the amplitude of the two modes has been written such that the
symmetry $\gamma \to 1-\gamma $ is manifest. Let us define the
dimensionless function ${\cal F}(\delta )$ by the following expression
\begin{equation}
\label{eq:defF}
{\cal F}(\delta)\equiv \frac{\pi}{2^{2\delta +1}\Gamma ^2(\delta
 +1/2)\cos^2(\pi \delta)}\, , 
\end{equation}
with $\delta =\gamma $ if $\gamma \le 1/2$ and $\delta =1-\gamma $ if
$\gamma \ge 1/2$. This function is represented in
Fig.~\ref{fig:amplitude}. If one assumes that the background space-time
is almost de Sitter, \ie $\beta \simeq -2$, as indicated by the fact
that the CMB power spectrum is almost scale-invariant, then one has
$\gamma \simeq -\alpha$. This function blows up at $\alpha =-1/2$ which
marks the frontier between the two branches. Otherwise, ${\cal F}(\alpha
)$ is smooth everywhere. In particular, the divergences due to the
cosine at the denominator are exactly cancelled out by the singularities
of the Euler function. For instance, if one considers the branch $\alpha
>-1/2$, this is best seen with the help of the equation, see
Ref.~\cite{Gradshteyn:1965aa}, $\Gamma\left(1/2+\delta \right)\cos
\left(\pi \delta \right)=\Gamma \left(1/2-\alpha\right)\cos \left(\pi
\alpha \right)=\pi/\Gamma \left(1/2+\alpha \right)$. At $\alpha =n/2$
with $n\in \setN $, the cosine vanishes but $\Gamma\left(1/2-\alpha
\right)=\Gamma(1/2-n/2)$ blows up since the Euler function is singular
for negative integer values of its argument. As proved by the above
formula, the net result is finite since $\Gamma\left(1/2+n/2\right)$ is
regular.

Finally, the Fourier energy density of the magnetic field (in other
words, the energy density stored at a given scale $k$) can be
expressed as
\begin{equation}
\label{eq:resultrhoB}
\frac{{\rm d}}{{\rm d}k}\rho_{_{B}}(\eta ,k)=\frac{k^3}{2\pi ^2}
{\cal F}(\delta)\frac{1}{a^4}\left(\frac{k}{aH}\right)^{2\delta} \, ,
\end{equation}
where, again, $\delta =\gamma $ if $\gamma \le 1/2$ and $\delta
=1-\gamma $ if $\gamma \ge 1/2$.

\subsection{The Coupling Function}
\label{subsec:couplingfunction}

In the previous section, we have derived the power spectrum of the
magnetic field under the assumption that $f\propto a^{\alpha }$. The
goal of this section is to obtain the explicit form of $f(\phi)$ for
various single field models of inflation.

\par

Let us start with power-law inflation since this is a situation where
an exact solution is available. In this case, the inflaton potential
is given by
\begin{equation}
\label{eq:potpowerlaw}
V(\phi)=V_0\exp\left[-\sqrt{2\epsilon _1}
\left(\phi -\phi_0\right)\right]\, ,
\end{equation}
where $\epsilon _1$ is the first slow-roll parameter and is
constant. Here, $\phi$ is dimensionless and denotes the vacuum
expectation value measured is units of $M_G$. Then, in conformal time,
the explicit solution of the Einstein equations reads
\begin{equation}
a(\eta )=\ell _0\left\vert \eta \right \vert ^{1+\beta }\, , \quad 
\phi (\eta )=\phi _0+\sqrt{2\epsilon_1}\left(1+\beta
\right)\ln \vert \eta \vert \, ,
\end{equation}
with $\beta \le -2$ and $\epsilon _1=(2+\beta )/(1+\beta)$. Therefore,
we see that, in order to satisfy our requirement $f\propto
a^{\alpha}$, we must choose the coupling function such that
\begin{equation}
\label{eq:fpowerlaw}
f(\phi)\propto \exp\left[\frac{\alpha} 
{\sqrt{2\epsilon _1}}
(\phi -\phi _0)\right]\, .
\end{equation}
This is a priori interesting since this is precisely the shape
postulated by Ratra in Ref.~\cite{Ratra:1991bn}. We will examine in
Sec.~\ref{sec:particlemodels} whether we can design a well-motivated
model which naturally gives this case.

\par

Let us now consider the case of a general potential. In this case, no
exact solution is available but the slow-roll approximation can be
used. Indeed, combining the two slow-roll equations of motion
\begin{equation}
H^2\simeq \frac{1}{3}V(\phi)\, , \quad \frac{{\rm d}\phi}{{\rm d}t}
\simeq -\frac{1}{3H}\frac{{\rm d}V}{{\rm d}\phi}\, ,
\end{equation}
one obtains 
\begin{equation}
\frac{{\rm d}a}{a}=- V(\phi)
\left(\frac{{\rm d}V}{{\rm d}\phi}\right)^{-1}{\rm d}\phi
\end{equation}
The integration of this equation is possible by quadrature. This
provides us with the function $a(\phi)$ from which we trivially
determine the form of the coupling function. One obtains
\begin{equation}
f(\phi )\propto \exp\left[-\alpha  \int ^{\phi }
\frac{V(\varphi )}{V'(\varphi )}
{\rm d}\varphi
\right]\, .
\end{equation}
The above expression gives the form of the coupling function for any
slow-roll model in terms of a single quadrature. In the following, we
apply this general formula to various slow-roll scenarios. Let us start
with the case of large field models, namely
\begin{equation}
\label{eq:largepot}
V(\phi )=M^4\phi^p \, ,
\end{equation}
for which we find the following coupling function
\begin{equation}
\label{eq:couplinglargefield}
f(\phi)\propto \exp\left(-\frac{\alpha }{2p}\phi ^2\right)\, .
\end{equation}
Then, we can also consider the case of small-field and hybrid inflation
where the potential can be written as
\begin{equation}
\label{eq:potsmallhybrid}
V(\phi )=M^4(1\pm\lambda\phi^p)
\end{equation}
where the (upper) plus sign corresponds to the hybrid case ($p=2$
being the theoretically favored value) while the (lower) minus sign
gives the small field case, and $\lambda$ is a constant.  The
corresponding coupling function reads
\begin{equation}
f(\phi)\propto \exp\left[\mp\frac{\alpha\phi^{2-p}}{\lambda p(2-p)}
-\frac{\alpha\phi^2 }{2p}\right]\, .
\end{equation} 
The case $p=2$ must be treated separately. One obtains
\begin{equation}
f(\phi)\propto
\phi^{\mp \alpha/(2\lambda)}\exp \left(-\frac{\alpha\phi^2}{4}\right)\, .
\end{equation} 
The case of small field is particularly interesting. Indeed, in this
case, $\phi\ll \mpl$, and the exponential term can be neglected. As a
result, one finds
\begin{equation}
\label{eq:couplingnew}
f(\phi)\propto \phi^{\alpha/(2\lambda)}\, .
\end{equation} 
The question is now to see whether one can find high-energy particle
physics models which reproduce the previous phenomenological
approach. This will be studied in detail in
Sec.~\ref{sec:particlemodels}.

\subsection{The Magnetic Field during preheating}
\label{subsec:preheating}

In this subsection, we study the behavior of the magnetic field during
the preheating stage. Previously, we have determined the shape of the
magnetic power spectrum and, in the following, we will evolve it until
present time using a simple adiabatic law. However, one must first check
that preheating is not going to affect either the amplitude or even the
spectral shape of ${\rm d}\rho _{_{B}}/{\rm d}k$. During preheating, at
least if the potential can be approximated by a parabola, the inflaton
field behaves according to
\begin{equation}
\label{eq:phipreheating}
\phi(t)\simeq \frac{\sqrt{8}}{\sqrt{3}mt}\sin \left(mt\right)\, .
\end{equation}
In the following we work in terms of cosmic time since this is more
convenient for our purpose. Then, the Fourier amplitude $A(t,k)$ obeys
the equation
\begin{equation}
\ddot{A}(t,k)+\left(3H+2\frac{\dot{f}}{f}\right)\dot{A}(t,k)
+\left(\frac{k^2}{a^2}+H^2+2H\frac{\dot{f}}{f}
+\frac{\ddot{a}}{a}\right)A(t,k)=0\, .
\end{equation}
Let us now consider the new rescaled variable $\aleph (t,k)$ defined by
$A(t,k)=\aleph (t,k)a^{-3/2}(t)f^{-1}(t)$. We are led to introduce a new
variable, slightly different from the one used before ${\cal A}(\eta
,k)$, because we now work in terms of cosmic time rather than in
conformal time. Using the above equation, it is straightforward to show
that
\begin{equation}
\label{eq:aleph}
\ddot{\aleph}(t,k)+\left(\frac{k^2}{a^2}-\frac{\ddot{f}}{f}
-H\frac{\dot{f}}{f}+\frac{1}{4}H^2-\frac12
\frac{\ddot{a}}{a}\right)\aleph(t,k)=0\, .
\end{equation}
In order to go further, one needs to know the time-behavior of $f(t)$
during preheating which requires the knowledge of the shape of the
inflaton potential during this stage of evolution. In the simple
approach considered here, this is the case only for chaotic
inflation. Indeed, the shape of $V(\phi)$ used in the power-law
case~(\ref{eq:potpowerlaw}) or in the small field/hybrid
case~(\ref{eq:potsmallhybrid}) is only valid during the slow-roll phase.
Therefore, let us now concentrate on the case of chaotic inflation. In
this case, using Eqs.~(\ref{eq:couplinglargefield})
and~(\ref{eq:phipreheating}) the coupling function is given by
\begin{equation}
f(t)\propto \exp\left[-
\frac{4\alpha}{3p (mt)^2}\sin ^2\left( mt \right) \right] \, .
\end{equation}
From this expression, one can evaluate the time-dependence of the
effective frequency in Eq.~(\ref{eq:aleph}). For the term $\ddot{f}/f$,
straightforward calculations lead to
\begin{eqnarray}
\frac{\ddot{f}}{f}&=&\frac{4\alpha}{3pt^2}
\left[-\frac{6}{(mt)^2}\sin ^2(mt)+\frac{3}{mt}\sin (2mt)-2\cos (mt)
\right]\nonumber \\
& +& \frac{16\alpha^2}{9p^2m^2t^4}
\left[\frac{4}{(mt)^2}\sin ^4(mt)-\frac{4}{mt}\sin ^3(mt)
+\sin ^2(2mt)\right]\, ,\\
&=& -\frac{8\alpha}{3pt^2}\cos (2mt)
+{\cal O}\left(\frac{1}{mt^3}\right)\, .
\end{eqnarray}
As a consequence, Eq.~(\ref{eq:aleph}) can be approximated as
\begin{equation}
\ddot{\aleph}(t,k)+\left[\frac{k^2}{a^2}+\frac{1}{4}H^2
+\frac{8\alpha}{3p t^2}\cos (2mt)
\right]\aleph(t,k)=0\, ,
\label{eq:aleph2}
\end{equation}
where we have retained the most and the least rapidly changing terms in
the bracket in order to compare with the Mathieu equation, ${\rm
d}^2\aleph/{\rm d}z^2+\left(a_{_{\rm M}}-2q_{_{\rm M}}
\cos2z\right)=0$. Then, defining $z\equiv mt-\pi/2$,
Eq.~(\ref{eq:aleph2}) is rewritten as
\begin{equation}
\frac{{\rm d}^2\aleph}{{\rm d}z^2}+\lkk \frac{k^2}{a^2m^2}
+\frac{H^2}{4m^2}-\frac{8\alpha}{3p(z+\pi/2)^2}\cos2z\rkk\aleph =0\, .  
\end{equation}
So we find 
\begin{equation}
a_{_{\rm M}}=\frac{k^2}{a^2m^2}+\frac{H^2}{4m^2}\, , \quad 
q_{_{\rm M}}=\frac{4\alpha}{3p(z+\pi/2)^2}\, ,
\end{equation} 
for a time scale much shorter than the Hubble time.  Since we are
interested in the long wave modes and $H \ll m$, we find $a_{_{\rm M}}
\ll 1$.  Then, the Mathieu equation has instability only for $q_{_{\rm
M}} \gtrsim 1$.  Since both $\alpha$ and $p$ are constants of order
unity with their typical values $\alpha=p=2$, we can conclude that the
instability continues for less than one period of the field
oscillation. The parametric resonance is therefore entirely negligible
and we cannot expect any enhancement of the magnetic field during
preheating, at least with the coupling function considered in this
article.

\subsection{The Magnetic Field after Inflation}
\label{subsec:afterinflation}

After inflation and preheating, the Universe is full of charged
particles and, therefore, the conductivity $\sigma_{\rm c}$ jumps up
to a value much larger than the Hubble parameter almost abruptly. In
this case, the model is described by the following action
\begin{equation}
\label{eq:actioncurrent} 
S\left[A_{\mu }\right]=-\int {\rm d}^4x \sqrt{-g}
\left(\frac{1}{4} g^{\alpha \beta }g^{\mu \nu}
F_{\mu \alpha }F_{\nu \beta }-j^{\mu }A_{\mu}\right)\, ,
\end{equation}
where, now, the function $f=1$ since the inflaton field has decayed
and is no longer present. The current $j^{\mu}$ is defined by the
formula~\cite{Barrow:2006ch}
\begin{equation}
j_{\mu}=\rho _{\rm e}u_{\mu}+\sigma _{\rm c}E_{\mu}\, .
\end{equation}
In this expression, $\rho _{\rm e}$ is the measurable charge density
and $\sigma _{\rm c}$ is the scalar conductivity of the medium. The
equation of motion now reads
\begin{equation}
-\frac{1}{\sqrt{-g}}\partial _{\mu}\left(\sqrt{-g}g^{\mu \nu}g^{\alpha \beta
}F_{\nu \beta }\right)=j^{\alpha}\, ,
\end{equation}
and the new stress-energy tensor can be expressed as 
\begin{equation}
T_{\mu \nu}=-g^{\alpha \beta }F_{\mu \alpha }F_{\beta \nu}
-\frac{1}{4}g_{\mu \nu}g^{\alpha \beta }g^{\gamma \delta }F_{\beta
\delta }F_{\alpha \gamma}+g_{\mu \nu}j_{\alpha }A^{\alpha}-2j_{\mu}A_{\nu}\, .
\end{equation}
Then, the wave equation for the gauge field reads
\begin{eqnarray}
\label{eq:eqwithconductivity}
\ddot{A_i}(t,\Vec{x}) + 
\left( \frac{\dot{a}}{a} + {\sigma}_\mathrm{c} \right)
        \dot{A_i}(t,\Vec{x})
 - {\partial}^j{\partial}_jA_i(t,\Vec{x}) = 0\, , 
\end{eqnarray}
In the large scale limit (hence neglecting the spatial derivatives of
the gauge field) and in the large conductivity limit, $\sigma_{\rm c}
\gg H$, the solution to the above equation reads
\begin{equation}
A_i(t,\Vec{x})=\frac{D_{1\, i}(\Vec{x})}{\sigma _{\rm c}}
{\rm e}^{-\sigma _{\rm
 c}t}+D_{2\, i}(\Vec{x})\, ,
\end{equation}
The exponential term will die away very quickly (with a characteristic
time $\tau =\sigma _{\rm c}^{-1}\gg 1/H$) and, therefore, one obtains
$A_i(t,\Vec{x})\simeq D_{2\, i}(\Vec{x})$. This implies that $E_i=0$ and
$B_i$ is a constant in time. Let us now evaluate the energy density
associated with the previous configuration. It is easy to see that the
extra term coming from the current is given by $ \rho =\sigma _{\rm c}
A^i\dot{A}_i$ and therefore vanishes. At the end, only the magnetic part
remains and reads
\begin{equation}
\rho _{_{\rm B}}=\frac{1}{4}\frac{1}{a^4}\delta ^{ij}\delta ^{k \ell}
(\partial _jD_{2\, \ell }-\partial _{\ell }D_{2\, j})
(\partial _iD_{2\, k}-\partial _{k}D_{2\, i})\propto \frac{1}{a^4}\, .
\end{equation}
Therefore, after inflation and until present times, one expects $\rho
_{_{B}}$ to scale as $1/a^4$ (independently of the era considered, ie
radiation or matter)\footnote{On large scales, one could worry that the
Ohm's law is not applicable and, hence, the previous result not
valid. However, even in this case, the conclusion that $\rho_{_{B}}$
scales as $1/a^4$ would be unchanged. Indeed, if the magnetic field is
described by Eq.~(\ref{eq:action}) but now with $f=1$ since the inflaton
field has decayed, then ${\cal A}={\rm e}^{-ik\eta }/\sqrt{2k}$ and
\begin{equation}
\rho _{_{B}}(\eta)=\frac{1}{2\pi ^2}\frac{1}{a^4}
\int_{0}^{+\infty}{\rm d}k k^3 \, .
\end{equation}
This integral diverges as $M_{_{\rm C}}^4$, where $M_{_{\rm C}}$ is a
cut-off as is usual for a vacuum contribution. But, as mentioned before,
the important point is that it still scales as $1/a^4$.}. As a
consequence, one deduces that the present magnetic energy density at a
given scale $L$ can be expressed as
\begin{equation}
\rho _{_{B}}(z=0, L)
=\frac{{\rm d}\rho _{_{B}}}{{\rm d}k}
\left(z=z_{\rm end}, k=\frac{2\pi}{L}\right)\frac{2\pi }{L}
\left(\frac{a_{\rm end}}{a_0}\right)^4\, .
\end{equation}
Using the expression of $\rho _{_{B}}$ at the end of inflation, one
obtains, with the definition $\Omega _{_{B}}(k)\equiv \rho
_{_{B}}(z=0,k)/\rho _{\rm cri}$,
\begin{equation}
\frac{{\rm d} \Omega _{_{B}}}{{\rm d}\ln k}= \frac{32}{9}{\cal
F}(\delta) \left(\frac{\rho _{\rm cri}}{\mpl ^4}\right)
\left(\frac{\rho _{\rm cri}}{\rho _{\rm end}}\right)^{\delta}
\left(\frac{a_0}{a_{\rm end}}\right)^{2\delta}
\left(\frac{k}{a_0H_0}\right)^{2\delta+4}
\end{equation}
To go further, one must therefore evaluate the ratio $a_0/a_{\rm
end}$. Clearly, this ratio depends on all the history of the Universe,
in particular, on the process on reheating. It has been shown in
Refs.~\cite{Martin:2006rs,Lorenz:2007ze} that it can be expressed as
\begin{equation}
\label{eq:a0aend}
\frac{a_0}{a_{\rm end}}=\frac{1}{R}\rho _{\rm end}^{1/2}
\left(\Omega _{\rm rad}^0\right)^{-1/4}\left(3
 H_0^2\right)^{-1/4}\, ,
\end{equation}
where $H_0\simeq 2.1 h\times 10^{-42}\mbox{GeV}$, $\Omega _{\rm
rad}^0h^2\simeq 4.3\times 10^{-5}$, see Ref.~\cite{Kolb:1990aa}, and
the parameter $R$, which describes the reheating phase, is given
by~\cite{Martin:2006rs,Lorenz:2007ze}
\begin{eqnarray}
\label{eq:lnR}
\ln R &\equiv & \frac{1-3w_{\rm reh}}{3(1+w_{\rm reh})}
\Biggl[\ln \left(g_*^{1/4}T_{\rm reh}\right)
+\frac{1+3w_{\rm reh}}{2(1-3w_{\rm reh})}\ln 
\rho _{\rm end} \nonumber \\
& & -\frac{1}{1-3w_{\rm reh}}\ln \left(
\frac{3+3w_{\rm reh}}{5-3w_{\rm reh}}\right)-\ln \frac{30^{1/4}}{\sqrt{\pi}}
\Biggr]\, .
\end{eqnarray}
This parameter depends, a priori, on three quantities: the reheating
temperature, $T_{\rm reh}$, the equation of state during reheating,
$w_{\rm reh}$, and the energy density at the end of slow-roll
inflation, $\rho _{\rm end}$. However, if one considers a particular
model then the number of free parameters can be reduced. For instance,
in the case of large field models where $V(\phi)\propto \phi ^p$, the
equation of state during reheating is known, namely
\begin{equation}
w_{\rm reh}=\frac{p-2}{p+2}\, .
\end{equation}
This leads to~\cite{Martin:2006rs,Lorenz:2007ze}
\begin{eqnarray}
\ln R &=& -\frac{p-4}{3p}
\Biggl[\ln \left(g_*^{1/4}T_{\rm reh}\right)
-\frac{p-1}{p-4}\ln \rho _{\rm end}\nonumber \\
& & +\frac{p+2}{2(p-4)}\ln \left(
\frac{3p}{p+8}\right)-\ln \frac{30^{1/4}}{\sqrt{\pi}}
\Biggr]\, .
\end{eqnarray}
Moreover, the value of $\rho _{\rm end}$ can also be
estimated. Indeed, the end of inflation is defined by the condition
\begin{equation}
\epsilon _1=-\frac{\dot{H}}{H^2}=\frac{3\dot{\phi }_{\rm end}^2/2}
{\dot{\phi}_{\rm end}^2/2+V(\phi_{\rm end})}=1\, ,
\end{equation}
which implies that $\dot{\phi }_{\rm end}^2=V(\phi _{\rm end})$ or
$\rho _{\rm end}=3V(\phi _{\rm end})/2$. Since $\phi _{\rm
end}/\mpl=1/(2\sqrt{\pi})$ for the potential $V(\phi)=m^2\phi ^2/2$,
one obtains $\rho _{\rm end }\simeq 3m^2\mpl^2/(16\pi)$, the mass $m$
being known, thanks to the COBE/WMAP normalization. Straightforward
calculations lead to $\rho _{\rm end}\sim 2160 \pi^2 /(2N_*+1)^2\times
Q_{\rm rms-PS}^2/T^2\sim 1.03 \times 10^{-11}$ with $Q_{\rm rms-PS}
\simeq 6\times 10^{-6}$, $T\sim 2.7\mbox{K}$ and assuming that the
number of e-folds between Hubble exit during inflation and the end of
inflation is given by $N_*=50$. We conclude that, in the case of a
massive large field model, the parameter $\ln R$ only depends on the
reheating temperature.

\par

The situation can be more complicated for other potentials. For small
field models, the shape of the potential during the slow-roll phase is
different from the shape of the potential during reheating. As a
consequence, the equation of state $w_{\rm reh}$ cannot be computed
from the parameters describing $V(\phi )$ as it was the case for large
field models. It remains a free parameter. In the case of power law
inflation and of hybrid inflation, one needs a mechanism of instability
in order to stop inflation. Then, even $\rho _{\rm end}$ remains a
free parameter since it is linked to the value of the inflaton field
where the waterfall behavior starts. Reheating proceeds in a
direction perpendicular (in the field space) to the inflationary
valley and, therefore, requires the full set ($w_{\rm reh}$, $T_{\rm
reh}$, $\rho _{\rm end}$) to be phenomenologically described.

\par

Putting everything together one arrives at the following expression
\begin{eqnarray}
\label{eq:ps}
\frac{{\rm d}\Omega _{_{B}}}{{\rm d}\ln k}
&=&\frac{{\cal F}(\delta )}{18\pi ^2}R^{-2\delta}
(\Omega _{\rm rad}^0)^{-\delta /2}
\rho _{\rm cri}^{1+\delta/2}
\left(\frac{k}{a_0H_0}\right)^{2\delta+4 }\, , \\
&\simeq & 2.4\times 10^{-7}\left(2.28\times 10^{-58}\right)^{\delta+2}
{\cal F}(\delta )R^{-2\delta}h^{2\delta +2}
\left(\frac{k}{a_0H_0}\right)^{2\delta +4}\, .
\end{eqnarray}
This equation is one of the most important result of this article. It
gives the amplitude, at a given scale, of the magnetic field today in
terms of four parameters: $\delta$, $\rho _{\rm end}$, $T_{\rm reh}$
and $w_{\rm reh}$.

\begin{figure}
\begin{center}
\includegraphics[width=12.5cm]{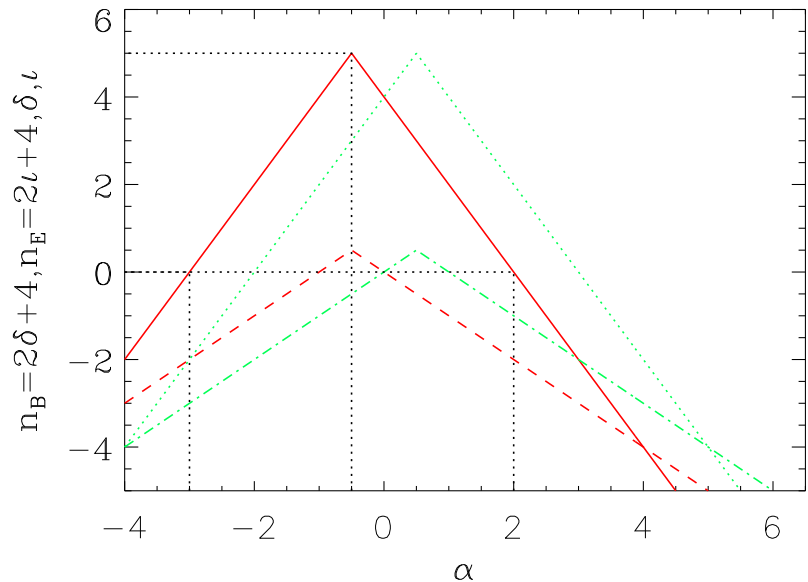} \caption{Spectral index
$n_{_{B}}$ (solid red line) of the magnetic power spectrum as a function
of the index $\alpha $, assuming, as before, a background expansion
closed to de Sitter, \ie $\beta \simeq -2$. Scale invariance of the
magnetic power spectrum corresponds to the values $\alpha =2 $ and
$\alpha =-3$. The dashed red line represents the quantity $\delta $ (see
the text) as a function of $\alpha $. The corresponding quantities for
the electric field are also displayed, namely the spectral index
$n_{_{E}}$ (green dotted line) and the function $\iota $ (green
dotted-dashed line), see the definition of $\iota $ in the text after
Eq.~(\ref{eq:defG}) and Eq.~(\ref{eq:psE}). Scale-invariance of the
magnetic power spectrum corresponds to a spectral index of the electric
power spectrum given by $n_{_{E}}=-2$ for $\alpha =-3$ and $n_{_{E}}=2$
for $\alpha =2$. Scale invariance of the electric power spectrum is
realized for $\alpha =-2$ or $\alpha =3$.}  \label{fig:tilt}
\end{center}
\end{figure}

In Fig.~\ref{fig:tilt}, we have represented the spectral index of the
magnetic power spectrum, $n_{_{B}}\equiv 2\delta +4$ as a function of
$\alpha$ as well as the quantity $\delta $ (see the red dashed
line). The two branches of the spectrum are easily identified and meet
at $\alpha =-1/2$. Scale invariance is obtained for $\alpha =2$ or
$\alpha =-3$. For values of $\alpha $ such that $-3<\alpha <2$, we have
a positive index, that is to say a blue spectrum while for $\alpha <-3$
and $\alpha >2$, we have a negative index \ie a red spectrum.

\par

In the next section, we find the constraints put on $\alpha $ by the
currently available data on the magnetic power spectrum.

\section{Constraining the coupling}
\label{sec:couplingconstraint}

\subsection{Limits on Cosmological Magnetic Fields}
\label{subsec:limits}

Before comparing the above results with observation we summarize upper
limits on cosmological magnetic fields from the following sources (see
more detailed explanations in
Refs.~\cite{Widrow:2002ud,Kolatt:1998}). The first type of constraints
comes from CMB anisotropy measurements. Indeed, homogeneous magnetic
fields during the time of decoupling whose scales are larger than the
horizon at that time cause the Universe to expand at different rates in
different directions.  Since anisotropic expansion of this type distorts
CMB, measurements of CMB angular power spectrum impose limits on the
cosmological magnetic fields.  Barrow, Ferreira, and
Silk~\cite{Barrow:1997mj} carried out a statistical analysis based on
the 4-year COBE data for angular anisotropy and derived the following
limit on the primordial magnetic fields that are coherent on scale
larger than the present horizon.
\begin{eqnarray}
\label{eq:cmbconslarge}
B_{\mathrm{cosmic}}^{(0)}<  5 \times10^{-9} \mathrm{G}\, .
\end{eqnarray}
Moreover, in Ref.~\cite{Jedamzik:1999bm}, a limit on primordial
small-scale magnetic fields was obtained, also from CMB
distortions. The constraint reads
\begin{eqnarray}
\label{eq:cmbconssmall}
B_{\mathrm{cosmic}}^{(0)}<  3 \times10^{-8} \mathrm{G}\, ,
\end{eqnarray}
for coherence lengths from $\sim 400 \mathrm{pc}$ and $\sim
0.6\mathrm{Mpc}$.

\par

Another type of constraint comes from Big Bang Nucleosynthesis (BBN)
since magnetic fields that existed during the BBN epoch would affect
the expansion rate, reaction rates, and electron density.  Taking all
these effects into account in calculation of the element abundances,
and then comparing the results with the observed abundances, one can
set limits on the strength of the magnetic fields.  The limits on
homogeneous magnetic fields on the BBN horizon size $\sim 1.4 \times
10^{-4}\mathrm{Mpc}$ are such that
\begin{equation}
\label{eq:bconsbbn}
B_{\mathrm{cosmic}}^{(0)}< 10^{-6} \mathrm{G}\, ,
\end{equation}
in terms of today's values~\cite{Grasso:1996kk,Cheng:1996yi}.

\par

Rotation Measure (RM) observations also provide interesting limits. RM
data for high-redshift sources can be used to constrain the large-scale
magnetic fields.  For example,\
Vall$\acute{\mathrm{e}}$e~\cite{valle:1990} tested for an RM dipole in a
sample of 309 galaxies and quasars.  The galaxies in this sample
extended to $z \simeq 3.6$ though most of the objects were at $z
\lesssim 2$.  Vall$\acute{\mathrm{e}}$e derived an upper limit of
$6\times10^{-10} {\left( {n_{e}}_0 /10^{-7}\mathrm{cm^{-3}}
\right)}^{-1}\mathrm{G}$, where ${n_{e}}_0$ is the present mean density
of thermal electrons, on the strength of uniform component of a
cosmological magnetic field ($ {n_{e}}_0 < {n_{b}}_0 = (2.7\pm0.1)
\times 10^{-7} \mathrm{cm^{-3}}$, where ${n_{b}}_0$ is the current
density of baryons~\cite{Spergel:2003cb}). Let us also notice that
Ref.~\cite{Blasi:1999hu} also quotes limits obtained from Faraday
measurements, namely
\begin{equation}
\label{eq:bconsfaraday50}
B_{\mathrm{cosmic}}^{(0)}< 6\times 10^{-9} \mathrm{G}\, ,
\end{equation}
for scales $\sim 50 \mathrm{Mpc}$ and 
\begin{equation}
\label{eq:bconsfaraday}
B_{\mathrm{cosmic}}^{(0)}< 10^{-8} \mathrm{G}\, ,
\end{equation}
for coherence lengths corresponding to $\sim \mathrm{G}$.

\par

Finally, one can also take into account a constraint which is of
different nature. It is known~\cite{Widrow:2002ud} that the magnetic
field on scales $\sim 1 \mathrm{Mpc}$ must be
\begin{equation}
\label{eq:bconsdynamo}
B_{\mathrm{cosmic}}^{(0)}>10^{-22} \mathrm{G}\, ,
\end{equation}
in order to ignite efficiently the dynamo mechanism at the galactic
scales, as was also mentioned in Sec.~\ref{sec:introduction}. This
constraint is clearly different from the above ones since it comes from
a theoretical prejudice. However, it is interesting since this leads to
a lower limit rather than upper bounds as discussed before.

\par

\begin{figure}
\begin{center}
\includegraphics[width=12.5cm]{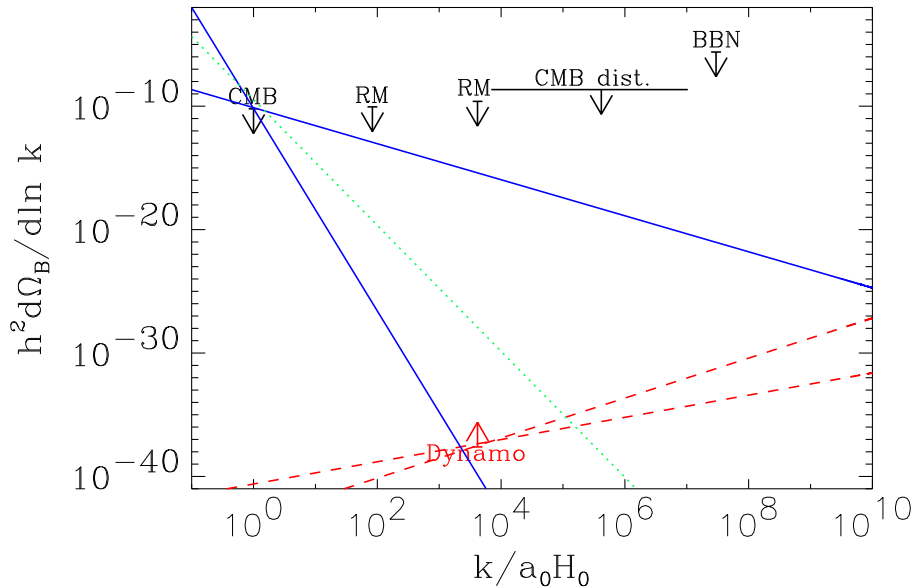} \caption{Constraints on the
amplitude of the present-day magnetic field at different scales. The
bounds come from CMB measurements, Big Bang nucleosynthesis (BBN) and
Faraday rotation (RM). The dynamo theoretical constraint is also
plotted. The two solid blue lines represent the reddest spectra
compatible with the CMB constraint at the Hubble scale in the case of a
general model of inflation and in the case of large field models,
respectively, while the dotted green line is the reddest spectrum
compatible with the CMB constraint and the dynamo limit (valid in the
case of a general model). Finally, the two dashed red lines are the
bluest spectra (for a general model and for large field, respectively)
compatible with the dynamo constraint.} \label{fig:bcons}
\end{center}
\end{figure}

These limits can be translated directly into limits on $\Omega
_{_{B}}$. Indeed, the energy density associated with the magnetic
field reads $\rho _{_{B}}=B^2/2$ (with our normalization of the gauge
field kinetic term, see also Ref.~\cite{Kolb:1990aa}) and this implies
${\rm d}\Omega _{_{B}}/{\rm d}\ln k=4\pi B^2/(3H_0^2\mpl^2)$.  Since
$1\mbox{T}^2/2\simeq 1.9\times 10^{-32}\mbox{GeV}^4$ and
$1\mbox{T}=10^4\mbox{G}$, one arrives at
\begin{equation}
h^2\frac{{\rm d} \Omega _{_{B}}}{{\rm d}\ln k}\simeq 2.425 \times 10^6
 \left(\frac{B}{1\mbox{G}}\right)^2\, .
\end{equation}
All the constraints on the present amplitude of the magnetic field are
summarized in Fig.~\ref{fig:bcons}.

\subsection{Constraining the Primordial Spectrum}
\label{subsec:consps}

In this section, our main goal is to constrain the parameter $\gamma$
using the experimental data reviewed in the previous
subsection. However, as already mentioned, the amplitude of the
theoretical spectrum given by Eq.~(\ref{eq:ps}) also depends on three
other parameters, namely $w_{\rm reh}$, $T_{\rm reh}$, $\rho _{\rm
end}$, that are not known precisely. Therefore, we first need to
discuss the constraints existing on the values of those parameters.
 
\par

Let us start with the limits on the parameters $R$. First of all,
there are limits on $\rho _{\rm end}$ since, in order to preserve the
success of the Big Bang Nucleosynthesis, one needs $\rho _{\rm
end}>\rho _{\rm nuc}\simeq 10^{-85}\mpl ^4$. On the other hand, one
knows that the energy scale of inflation is constrained from above,
namely $H_{\rm inf}/\mpl <1.3\times 10^{-5}$, see
Ref.~\cite{Martin:2006rs}. As a consequence, one must require that
$\rho _{\rm end}<10^{-10}\mpl ^4$. Summarizing the constraints, one has
\begin{equation}
-187 < \ln \rho _{\rm end}<-20 \, .
\end{equation}
Let us now consider the equation of state during reheating, $w_{\rm
reh}$. Reheating is, by definition, a phase of evolution where the
scale factor does not accelerate which means that the strong and
dominant energy conditions should be satisfied, that is to say
\begin{equation}
-\frac13<w_{\rm reh}<1\, .
\end{equation}
Finally, one also gets constraints on $\rho _{\rm reh}=g_*\pi
^2T^4_{\rm reh}/30$ from the obvious requirement that reheating must
proceed after the end of inflation and before the BBN, that is to say
\begin{equation}
\rho _{\rm nuc}<\rho_{\rm reh}<\rho_{\rm end}\, .
\end{equation}
Taking everything into account, one arrives at the following
model-independent range of variation
\begin{equation}
\frac14 \ln \rho _{\rm nuc}<\ln R< 
-\frac{1}{12} \ln \rho _{\rm nuc}
+\frac{1}{3}\ln \rho _{\rm end}\, ,
\end{equation}
up to negligible factors depending on $w_{\rm reh}$ only. This leads to
\begin{equation}
\label{eq:boundsonR}
-47\lesssim \ln R \lesssim 10 \, .
\end{equation}
In Ref.~\cite{Martin:2006rs}, CMB data have been used to constrain the
values of $\ln R$ for large, small, hybrid and running mass
inflationary scenarios. Some very weak limits have been obtained in
the case of small field models but, essentially, the currently
available data are not yet accurate enough and the marginalized
probabilities obtained are just cut at the edges of their
priors. Therefore, in absence of any other constraints, one must
consider the interval given by Eq.~(\ref{eq:boundsonR}). This is what
is assumed here for power-law inflation even if, strictly speaking,
the analysis on $\ln R$ should be redone for this model which was not
explicitly considered in Ref.~\cite{Martin:2006rs}. However, it seems
extremely likely that no strong constraints would be obtained with the
present-day data. In addition, let us recall that, in order to stop
inflation, one must implement a mechanism of tachyonic instability
where $\rho _{\rm end}$ is now viewed as a free parameter of the
model. In this context, the constraint of Eq.~(\ref{eq:boundsonR})
seems to be particularly relevant.

\begin{figure}
\begin{center}
\includegraphics[width=9.5cm]{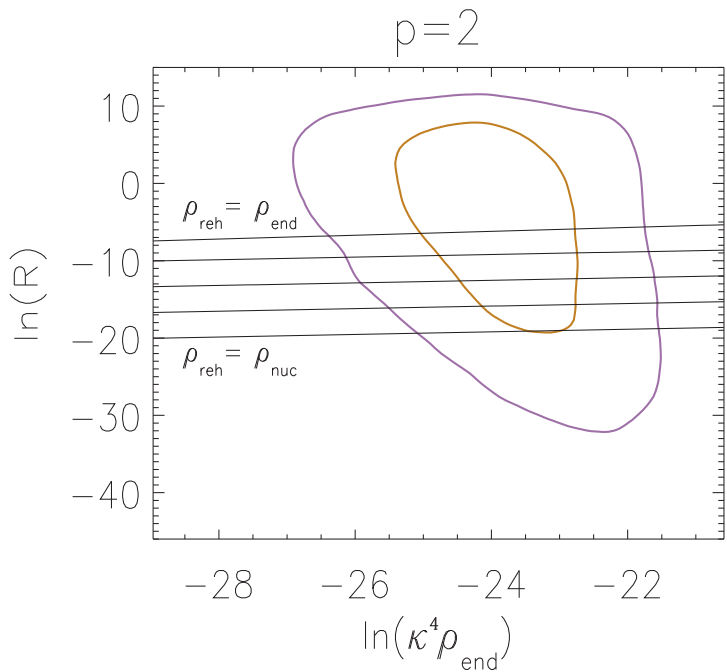} 
\caption{Constraints obtained from the WMAP3 data on the parameter
$\ln R$ at $1\sigma$ and $2\sigma$, see Ref.~\cite{Martin:2006rs}.
The solid lines represent the variation of $\ln R$ as a function of
the energy density at the end of inflation for various reheating
temperatures (or, equivalently, for various energy densities at the
end of reheating). No observational constraint on $T_{\rm reh}$ is
obtained in the case of large field models.}
\label{fig:lnR}
\end{center}
\end{figure}

As already mentioned, the situation is slightly different for the
massive large field model, $p=2$. Indeed, in this case, the equation
of state during reheating is known $w_{\rm reh}=0$, as well as $\rho
_{\rm end}$ (at least approximatively, see above). In this case, one
obtains the following constraint $-20\lesssim \ln R \lesssim -6$. In
Fig.~\ref{fig:lnR}, one has reproduced this range of variation
together with the constraints on $\ln R$ obtained by scanning general
models of the form, $V(\phi)\propto \phi ^p$, for more details see
Ref.~\cite{Martin:2006rs}. This plot confirms that the reheating
temperature remains a free quantity with the currently available CMB
data.

\par

Let us now turn to the constraints on $\alpha $. In the case where the
spectral index $n_{_{B}}$ is negative (red spectrum), the limiting
constraint is the CMB constraint at scales of the order of the horizon
today, see Fig.~\ref{fig:bcons}. So, one should require
\begin{equation}
\label{eq:inequalcmb}
\log _{10}\left(h^2\frac{{\rm d}\Omega _{_{B}}}{{\rm d}\ln k}\right)
<\log _{10}\left(h^2\frac{{\rm d}\Omega _{_{B}}}{{\rm d}\ln k}\Biggl
\vert _{_{\rm CMB}}\right)\simeq -10.2 \, .
\end{equation}
According to Fig.~\ref{fig:tilt}, $n_{_{B}}<0$ means either $\alpha
>2$ or $\alpha <-3$. Looking at the dashed red line, one sees that
this implies $\delta <0$ in these two regimes and for this reason, one
writes $\delta =-\vert \delta \vert$. Then, using Eq.~(\ref{eq:ps})
and working out Eq.~(\ref{eq:inequalcmb}), one obtains
\begin{eqnarray}
\vert \delta \vert < \frac{122.47
+\log _{10}\left(h^2{\rm d}\Omega _{_{B}}/{\rm d}\ln k
\vert _{_{\rm CMB}}\right)
-\log_{10}{\cal
F}-4\log_{10}\left[k/\left(a_0H_0\right)\right]}
{57.92+0.86\ln
R-2\log_{10}\left[k/\left(a_0H_0\right)\right]}\, ,
\nonumber \\
\end{eqnarray}
where we have used $h\simeq 0.72$. In fact, this inequality is not as
simple as it may seem since ${\cal F}$ is also a function of $\delta
$. Therefore, given the functional form of ${\cal F}$, only a
numerical calculation could allows us to derive the exact bound on
$\delta$. However, the impact of the term $\log _{10}({\cal F})$ is
limited and can be neglected in a first approximation. Moreover, since
we seek an upper limit, it is clear from the above expression that the
smallest value of $\ln R$ should be considered in order to reduce the
denominator. According to the considerations presented before, this
means $\ln R\sim -47$ in general and $\ln R \sim -26$ for large field
models. Then, using $k/(a_0H_0)=1$ and $\log _{10}\left(h^2{\rm
d}\Omega _{_{B}}/{\rm d}\ln k \vert _{_{\rm CMB}}\right)\simeq
-10.21$, one arrives at $\vert \delta \vert \lesssim 6.1$ and $\vert
\delta \vert \lesssim 2.72$ for large fields. The corresponding
spectra are represented in Fig.~\ref{fig:bcons}, see the two solid
blue lines. If $\alpha >2$, then $\delta =-\alpha $ and the two bounds
respectively correspond to $\alpha <6.1$ and $\alpha <2.72$. On the
other hand, if $\alpha <-3$, then $\delta =1+\alpha $. As a
consequence, one obtains $\alpha >-7.1$ and $\alpha >-3.72$. However,
one notices in Fig.~\ref{fig:bcons} that the spectra with $\alpha
=6.1$ and $\alpha =-7.1$ violate the dynamo limit (this is not the
case for large field models). Therefore, one should repeat the
calculation and require the dynamo limit to be also
satisfied. Straightforward calculations indicate that this amounts to
$\alpha <4.55$ and $\alpha >-5.55$ (with, this time, $\ln R\sim
-39.2$). The corresponding spectrum is represented in
Fig.~\ref{fig:bcons} by the dotted green line.

\par

Let us now consider the case where we have a blue spectrum,
$n_{_{B}}>0$. According to Fig.~\ref{fig:bcons}, this means $-3<\alpha
<2$. This time the situation could be more complicated because this
range of values can correspond to positive or negative values of
$\delta $, see the dashed red line in Fig.~\ref{fig:tilt}. However, in
practice, one remains in the regime where $\delta <0$ as will be
checked below and, hence, $\delta =-\vert \delta \vert$. For blue
spectra, the relevant constraint is the dynamo constraint which reads
\begin{equation}
\vert \delta \vert > \frac{78.23-\log_{10}{\cal
F}}{50.68+0.86\ln R}\, , 
\end{equation}
where we have used 
\begin{equation}
\log _{10}\left(h^2\frac{{\rm d}\Omega _{_{B}}}{{\rm d}\ln k}\Biggl
\vert _{_{\rm Dynamo}}\right)\simeq -37.61 \, ,
\end{equation}
and $\log _{10}[k/(a_0H_0)]\sim 3.62$, the scale at which the dynamo
constraint applies. Since we now seek a lower bound on $\vert \delta
\vert $, one should use the largest value of $\ln R$, that is to say
$\ln R\sim 10$ in general and $\ln R \sim -6$ for large field
models. This respectively gives $\vert \delta \vert >1.19$ and $\vert
\delta \vert >1.55$. This implies $\alpha >1.19$ and $\alpha >1.55$ in
the case where $\alpha >-1/2$ (and, hence, one verifies that, for
these values of $\alpha $, $\delta $ is indeed negative, see
Fig.~\ref{fig:tilt}). In the other case, this means $\delta =1+\alpha
<0$ and, therefore, one has $\alpha <-2.19$ and $\alpha <-2.55$. The
corresponding spectra are represented in Fig.~\ref{fig:bcons} by the
two dashed red lines, the one with the smallest slope being the
spectra in the case of large field models.

\par

Summarizing all the results obtained above, one obtains the following
general constraints
\begin{equation}
\label{eq:range}
-5.55 <\alpha <-2.19\, , \quad 1.19<\alpha <4.55 \, ,
\end{equation}
while, for the case of large field models, one arrives at
\begin{equation}
-3.72<\alpha <-2.55\, ,\quad 1.55<\alpha <2.73\, .
\end{equation}
This means that, if $\alpha $ is in the above ranges of values, then
there exists at least one value of $\ln R$ such that the corresponding
spectrum is compatible with the currently available data.
   
\subsection{Consistency and the Back-reaction Problem}
\label{subsec:consistency}

We now evaluate the electric field produced in the model under
investigation. Our goal in this subsection is to check that the total
amount of magnetic and electric energy density produced during
inflation is not larger than the background energy density $\sim
H_{\rm inf}^2\mpl^2$. Otherwise, clearly, the framework used here
would not be consistent and would suffer from a serious back-reaction
problem. Using the expression of the stress-energy tensor, see
Eq.~(\ref{eq:stresstensor}), one obtains for the electric time-time
component
\begin{equation}
T^E{}^0{}_0=-\frac12 \frac{f^2}{a^2}g^{ij}\partial _0A_i\partial _0A_j \, ,
\end{equation}
from which straightforward calculations lead to the following
expression for the vacuum energy density
\begin{equation}
\left \langle T^E{}^0{}_0\right \rangle =-\frac{1}{2\pi ^2}
\frac{f^2}{a^4}\int {\rm d}kk^2\left\vert \left(\frac{{\cal A}}{f}\right)'
\right\vert ^2\, ,
\end{equation}
or, in terms of electric energy density stored at the scale $k$ 
\begin{equation}
\frac{{\rm d}}{{\rm d}k}\rho _{_{E}}=\frac{f^2}{2\pi ^2}\frac{k^2}{a^4}
\left\vert \left[\frac{{\cal A}(k,\eta )}{f}\right]'
\right\vert ^2 \, .
\end{equation}
This equation should be compared to Eq.~(\ref{eq:rhoB}). The solution
for the Fourier amplitude of the vector potential has already been
obtained previously in terms of Bessel functions. Using
Eq.~(\ref{eq:sol}) on large scales compared to the Hubble radius, that
is to say $k/(a H) = -k\eta \to 0$, one obtains
\begin{eqnarray}
{\cal A} '-\frac{f'}{f}{\cal A} &\to& 
\frac{\sqrt{\pi}}{2^{\gamma+3/2}}
\frac{{\rm e}^{i\pi \gamma /2-i\pi}}{\Gamma (\gamma+3/2)
\cos (\pi\gamma )}k^{1/2}(k\eta )^{\gamma+1}
\nonumber \\
& & +\frac{\sqrt{\pi}}{2^{-\gamma +1/2}}
\frac{{\rm e}^{-i\pi (3+\gamma)/2}}{\Gamma (-\gamma+1/2)
\cos \left[-\pi(1+\gamma)\right]}k^{1/2}(k\eta )^{-\gamma}\, ,
\end{eqnarray}
where, as was already done for the magnetic field, the amplitude of
the two modes has been written such that the symmetry $\gamma \to
-1-\gamma $ is manifest. Clearly, this symmetry is not exactly similar
to the one obtained in the magnetic case. Let us now define the
dimensionless function ${\cal G}(\iota )$ by the following expression
\begin{equation}
\label{eq:defG}
{\cal G}(\iota)\equiv \frac{\pi}{2^{2\iota +3}\Gamma ^2(\iota
 +3/2)\cos^2(\pi \iota)}\, , 
\end{equation}
with $\iota =\gamma +1$ if $\gamma \le -1/2$ and $\iota =-\gamma $ if
$\gamma \ge -1/2$. This definition is very similar to the
definition~(\ref{eq:defF}). The functions $\iota (\alpha )$ and ${\cal
G}(\alpha)$ are represented in Figs.~\ref{fig:tilt}
and~\ref{fig:amplitude} respectively. With the help of ${\cal
G}(\iota)$, we can write the Fourier energy density of the electric
field. One obtains
\begin{equation}
\label{eq:psE}
\frac{{\rm d}}{{\rm d}k}\rho_{_{E}}(\eta ,k)=\frac{k^3}{2\pi ^2}
{\cal G}(\iota)\frac{1}{a^4}\left(\frac{k}{aH}\right)^{2\iota} \, .
\end{equation}
This expression should be compared to the formula expressing the
magnetic energy density at a given scale $k$, see
Eq.~(\ref{eq:resultrhoB}).

\par

We are now in a position where one can estimate when a back-reaction
problem occurs. Clearly, the model is free of this difficulty when the
following condition is satisfied
\begin{equation}
\frac{{\rm d}\rho _{_{E}}}{{\rm d}\ln k}\biggl \vert _{\rm inf}
+\frac{{\rm d}\rho _{_{B}}}{{\rm d}\ln k}\biggl \vert _{\rm inf}
<\rho _{\rm inf}\, ,
\end{equation}
where, in the present context, the subscript ``inf'' means evaluated at
the end of inflation. Indeed, as discussed previously, after the end of
inflation, the conductivity jumps and, as a consequence, the gauge field
becomes constant and, therefore, the electric field vanishes. Thus, if
one checks that, at the end of inflation, the electric field can not
cause a back-reaction problem, then we are guaranteed that the complete
scenario is consistent. Let us also notice that the previous equation
should be satisfied at any scales of astrophysical relevance
today. Using the expression of the magnetic and electric energy
densities derived before, the above relation amounts to
\begin{eqnarray}
\label{eq:consistency}
& & \frac{H_{\rm inf}^4}{2\pi ^2}\Biggl[{\cal
      G}(\iota)\left(\frac{a_0H_0}{a_{\rm inf}H_{\rm
      inf}}\right)^{2\iota +4}\left(\frac{k}{a_0H_0}\right)^{2\iota
      +4}\nonumber \\ & & +{\cal F}(\delta)\left(\frac{a_0H_0}{a_{\rm
      inf}H_{\rm inf}}\right)^{2\delta
      +4}\left(\frac{k}{a_0H_0}\right)^{2\delta
      +4}\Biggr]<\frac{3}{8\pi}H_{\rm inf}^2\mpl^2 \, .
\end{eqnarray}
In the following, for simplicity, we will not distinguish the energy
density at the end of inflation from the energy density at which the
scales of astrophysical interest today left the Hubble scale. Clearly,
this is a good approximation during inflation (almost by
definition). Under this assumption, using Eq.~(\ref{eq:a0aend}), one
can estimate the ratio $a_0H_0/(a_{\rm inf}H_{\rm inf})$ which enters
the constraint~(\ref{eq:consistency}). One obtains
\begin{equation}
\label{eq:approxratioaH}
\frac{a_0H_0}{a_{\rm inf}H_{\rm inf}}\simeq 1.51\times
10^{-29}\frac{h}{R}\, .
\end{equation}
Therefore, one sees that the constraint~(\ref{eq:consistency}) is
quite difficult to analyze in full generality since, as already
discussed, the quantity $R$ depends on three parameters, namely the
energy density at the end of inflation, the equation of state parameter
during reheating and the reheating temperature, see also
Eq.~(\ref{eq:lnR}). In the following, instead of performing a
systematic scanning of the parameter space, we just show that, in the
vicinity of the ``scale-invariant'' values $\alpha =-3$ and $\alpha
=2$, there exist consistent models of inflation where the
back-reaction problem does not occur. For this purpose, for
simplicity, one first assumes instantaneous reheating, $g_*T_{\rm
reh}^4 \sim 90H_{\rm end}^2\mpl^2/(8\pi ^3)$. Then, the expression of
$\ln R$ simplifies considerably and reads
\begin{equation}
\label{eq:approxlnR}
\ln R \simeq \frac{1}{4}\ln \rho _{\rm end}\, .
\end{equation}
Then it is straightforward to work out the
constraint~(\ref{eq:consistency}). For $\alpha =-3$, using
Eqs.~(\ref{eq:approxratioaH}) and~(\ref{eq:approxlnR}), one obtains
\begin{equation}
\frac{H_{\rm inf}}{\mpl}<{\cal O}(1)\times
10^{-20}\left(\frac{k}{a_0H_0}\right)^{2/3}\, .
\end{equation}
Since, in the present situation, the magnetic power spectrum is scale
invariant, the above constraint comes almost entirely from the
requirement that the electric energy density should be less that the
inflationary background energy density. Moreover, for $\alpha =-3$,
one has a red electric power spectrum with $n_{_{E}}=-2$, see
Fig.~\ref{fig:tilt}. As a consequence, if the above constraint is
satisfied at the Hubble scale, $k/(a_0H_0)\simeq 1$, then it is
satisfied at any scales of astrophysical relevance today. Therefore,
for the case $\alpha \sim -3$, one finds $H_{\rm inf}/\mpl \lesssim
10^{-20}$ or $\rho_{\rm inf}\sim 10^{-41}\mpl ^4$ (let us recall that
$\rho _{\rm nuc}\sim 10^{-85}\mpl ^4$).

\par

\begin{figure}
\begin{center}
\includegraphics[width=9.5cm]{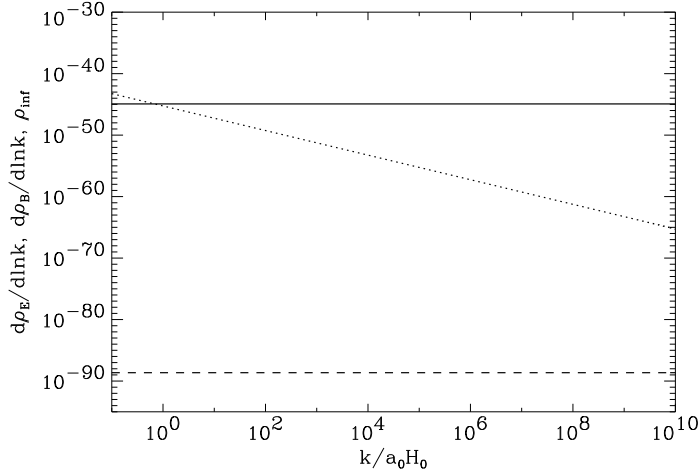} 
\caption{Inflationary (solid line), magnetic (dashed line) and
  electric (dotted line) energy densities at the end of inflation in
  the case where $\alpha =-3$, $H_{\rm inf}/\mpl =10^{-22}$
  $n_{_{B}}=0$ and $n_{_{E}}=-2$. This situation corresponds to the
  limiting case treated in the text where there is no back-reaction
  problem, the electric energy density being always below the
  background energy density and where the dynamo constraint is still
  satisfied.}
\label{fig:valid}
\end{center}
\end{figure}

However, we still have to check that the previous models satisfy the
constraints of figure~\ref{fig:bcons}. Indeed, we have shown before
that, for indices $\alpha $ in the ranges~(\ref{eq:range}), there is
always a value of $\ln R$, compatible with the CMB data, such that the
model is in agreement with the observations. However, we have not
proven that the models are compatible for any values of $\ln R$ (this
is clearly not the case) and, hence, since $\ln R$ is now fixed (for
instance with $w_{\rm reh}=0$ and $H_{\rm inf}\sim 10^{-20}\mpl$, one
has $\ln R\sim -21.77$) one must verify that there is no problem with
the data. In fact, it turns out that these models do not satisfy the
dynamo constraint because the magnetic field is too strongly damped
after inflation. If we want to solve this problem, one has to consider
a different type of reheating era. Clearly, the most favorable
situation is when $w_{\rm reh}\sim 1$ since this is the situation
where $B$ scales the more slowly given that $-1/3<w_{\rm reh}<1$,
namely $B\propto t^{-2/3}$ and when the reheating stage is prolonged
as much as possible, that is to say down to a reheating temperature of
the order of few $\mbox{MeV}$'s. In this case avoiding the
back-reaction problem requires $H_{\rm inf}<10^{-22}\mpl$ which leads
to $\ln R\sim -16.74$. The corresponding situation is illustrated in
Fig.~\ref{fig:valid}. This time, for the previous value of $\ln R$,
one can check that the branch $\alpha =-3$ of the spectrum is still
compatible with the dynamo constraint. Therefore, at least for this
type of reheating period, this branch is still a viable alternative.

\par

The case $\alpha \sim 2$ can be worked out in the same
manner. Straightforward calculations lead to
\begin{equation}
\frac{H_{\rm inf}}{\mpl}<{\cal O}(1)\times
10^{58}\left(\frac{k}{a_0H_0}\right)^{-2}\, .
\end{equation}
This time the electric spectral index is positive, $n_{_{E}}=2$ and,
therefore, the power spectrum is blue. As consequence, one should
check that the consistency relation~(\ref{eq:consistency}) is
satisfied at small scales, say $k/(a_0H_0)\sim 10^{10}$. As is obvious
from the above relation, this is very easily the case.

\par

The fact that back-reaction is always satisfied when $\alpha \sim 2$
while it requires low scale inflation in the case $\alpha \sim -3$ can
be easily understood, see also Ref.~\cite{Bamba:2003av}. Indeed, from
Eq.~(\ref{eq:FourierA}), one has $A_i(\eta ,\Vec{x})\sim \epsilon
_{i\lambda }A(\eta ,k)\sim {\cal A}(\eta , k)/f(\eta )$, where we have
used the fact that the covariant polarization vector contains a factor
$a(\eta)$, see Eq.~(\ref{eq:defepsilon}). Then, from
Eq.~(\ref{eq:fourieramplitude}), one has the two following modes on
larges scales ${\cal A}\propto f(\eta ) $ and ${\cal A}\propto f(\eta
)\int ^{\eta }{\rm d}\tau/f^2(\tau)$. Clearly, from the above equations,
the first mode leads to $A_i\sim \mbox{Const.}$ and, therefore, to a
vanishing electric field, while the second remains time-dependent and
can be responsible for the production of a strong electric
field. Indeed, it is easy to see that, in this last case, $A_i(\eta
,\Vec{x}) \propto a^{-1-2\alpha }$, that is to say $A_i(\eta
,\Vec{x})\propto a^{5}$ for $\alpha \sim -3$ and $A_i(\eta ,
\Vec{x})\propto a^{-5}$ for $\alpha \sim 2$. Therefore, this mode is
sub-dominant for $\alpha \sim 2$ but is dominant for the other branch of
the spectrum $\alpha \sim -3$. In this case, its strong time-dependence
$\propto a^{5}$ produces a strong electric field and this explains why
the back-reaction problem can be severe in this situation. This problem
can be avoided if the production of the electric field is limited as it
happens to be the case if inflation proceeds at low scale.

\section{Particle Physics Models}
\label{sec:particlemodels}

\subsection{Power Law Inflation}
\label{subsec:particlepl}

In this section, we finally turn to the model-building issue. More
precisely, the question addressed in this subsection is whether there is
a natural high energy model with an exponential inflaton potential and
an exponential gauge coupling, as was obtained in
Sec.~\ref{subsec:couplingfunction}. We will consider the other models in
the next subsections. In order to study this issue, let us consider that
the inflaton field is a modulus field $T$ in string theory. Then, its
K\"ahler and super potential are given by
\begin{equation}
K=-p\ln\left(T+T
^{\dagger}\right)\, , \quad W=-\frac{M^3}{n}T^n\, ,
\end{equation}
the natural choice being $p=3$ (no-scale). Moreover, in string theory, a
natural choice is $f(T)=T^q$ with the preferred choice
$q=1$~\cite{Bailin:1994}. Therefore, at first sight, this does not give
the required $f$ function. However, one has to remember that the field
$T$ is not canonically normalized because the K\"ahler potential is
non-trivial. The canonical field $\phi $ is given by
\begin{equation} 
T={\rm e}^{-\sqrt{2/p}\phi }\, .
\end{equation}
As a consequence, one obtains a coupling function with the required
exponential shape, namely
\begin{equation}
f(\phi )\propto {\rm e}^{-q\sqrt{2/p}\phi}\, .
\end{equation}
Moreover, the F-term potential is also easily computed and one also
obtains the required shape, namely an exponential dependence. The
explicit expression reads
\begin{equation}
\label{eq:potpl}
V\left(\phi \right)=\frac{
M^6(n-3)}{6n}\exp\left[-\left(2n-3\right)
\sqrt{\frac{2}{3}}\phi \right]\, .
\end{equation}
If we compare the two previous formulas to Eqs.~(\ref{eq:fpowerlaw})
and~(\ref{eq:potpowerlaw}), then one obtains the correct power-law model
for the choice
\begin{equation}
n=\frac32\left(1+\sqrt{\frac{\epsilon _1}{3}}\right)\, ,\quad 
q=-\frac{\alpha }{2}\sqrt{\frac{p}{\epsilon _1}}\, .
\end{equation}
In particular, we see that the branch $\alpha \sim 2$ leads to a
negative power for $q$ while the branch $\alpha \sim -3$ gives
$q=3\sqrt{3}/(2\sqrt{\epsilon _1})\gg 1$. This is already an indication
that it is difficult to reconcile the branch $\alpha \sim 2$ with a
sensible model building. In fact, in the present context, even the
$\alpha \sim -3$ branch cannot lead to the string inspired favored
model with $q=1$. But, the lethal flaw of the above model is that the
required value of $n$ is less than $3$ leading to a negative inflaton
potential.

\par

The next question is whether the mechanism of assisted
inflation~\cite{Liddle:1998jc} could solve this situation. Let us
envisage a situation where we have $N$ moduli such that
\begin{equation}
K=-\sum _{i=1}^N p_i\ln \left(T_i+T_i
^{\dagger}\right)\, , \quad W=-\sum
_{i=1}^N\frac{M^3}{n_i}T_i^{n_i}\, ,
\end{equation} 
the no-scale structure being preserved if $\sum _{i=1}^Np_i=3$. As
before, the field $T_i$ are not canonically normalized, the normalized
field $\phi _i$ being given by
\begin{equation} 
T_i={\rm e}^{-\sqrt{2/p_i}\phi _i}\, .
\end{equation}
Straightforward manipulations lead to the following potential
\begin{eqnarray}
V &=& \frac{ M^6}{2}\exp\left(\sum
_{i=1}^N\sqrt{2p_i}\phi _i\right)
\Biggl[-\sum _{j=1}^N\sum _{k=1}^N \frac{1}{n_k}
\exp\Biggl(-n_j \sqrt{\frac{2}{p_j}}\phi _j
\nonumber \\ & & -n_k \sqrt{\frac{2}{p_k}}
\phi _k\Biggr)+\sum _{j=1}^N\frac{1}{p_j}
\exp\left(-2n_j \sqrt{\frac{2}{p_j}}\phi _j
\right)
\Biggr]
\end{eqnarray}
If one considers the case of one field (which implies $p=3$), then one
can check that Eq.~(\ref{eq:potpl}) is correctly recovered. Then, let us
consider that we have $N$ fields such that $p_i=3/N$ and $n_i=n$ for any
index $i$. Then, the potential becomes
\begin{equation}
V(\phi)=\frac{
M^6N^2(n-3)}{6n}\exp\left[-\left(2n-3N\right)
\sqrt{\frac{2N}{3}}\phi \right]\, .
\end{equation}
This implies that
\begin{equation}
n=\frac{3N}{2}\left(1+\frac{1}{N^{3/2}}\sqrt{\frac{\epsilon
	       _1}{3}}\right)\, .
\end{equation}
We see that $N\ge 2$ allows us to have a positive potential and,
therefore, represents a meaningful model. Of course, the fact that the
power of the superpotential is not an integer is not a nice feature and,
for this reason, the model remains unsatisfactory. What about the gauge
coupling? In presence of several superfields, the form of $f$ that
should be chosen remains unclear. If we postulate that it depends on one
superfield only (say the field $\phi _i$), then one gets
\begin{equation}
q=-\frac{\alpha}{2}\sqrt{\frac{p_i}{\epsilon _1}}\, ,
\end{equation}
In this case, and contrary to the single field case, $q$ is not
necessarily very large (as one might think since it scales as the
inverse of the slow-roll parameter) because $p_i$ can be a small
number. In the case envisaged above, this is generically the case
because $p_i=3/N$ (remember that only the condition $\sum _{i=1}^N
p_i=3$ must be fulfilled in order to preserve the no-scale
structure). Moreover, one could also imagine a case where all the
coefficients $p_i$ are not equal. If one of the $p_i$ is relatively
small, then it could compensate the inverse of the slow-roll parameter
in order to give a number of order one. For the branch $\alpha \sim 2$,
$q$ is negative and we are not aware of any well-motivated SUGRA model
where this happens. Therefore, the most interesting case seems to use
the other branch of the spectrum with $\alpha \sim -3$. Then, one
obtains
\begin{equation}
q\sim \frac{\vert \alpha \vert}{2}\sqrt{\frac{3}{N\epsilon _1}}\, ,
\end{equation}
with $\vert \alpha \vert \sim 3$ and the favored model $q=1$ is
obtained for
\begin{equation}
N\sim \frac{3\alpha ^2}{4}\frac{1+\beta }{2+\beta }\, .
\end{equation}
Of course $N$ should be an integer and this can be the case if $\alpha $
is not exactly $-3$. We have seen that this situation is perfectly
compatible with the data, see the previous section. Let us give an
example for the sake of illustration. We can take $\beta \sim -2.025$
leading the spectral index $\nS=0.95$. Then, the value $\alpha
=-3.00135$ gives $N=277$ fields in the model.
 
\par

A last remark is in order here. We have seen that the branch $\alpha
\sim -3$ is a viable solution only if the scale of inflation is low and
the reheating stage quite long. Here, we can certainly design a model
where this is the case since, in order to stop inflation, one needs to
rely on a mechanism of tachyonic instability as in the case of hybrid
inflation. In this situation, the energy density at the end of inflation
remains a free parameter and can be chosen such that it leads to the
correct scale of inflation. Moreover, the details of the reheating will
depend of the shape of the potential in the direction perpendicular to
the inflationary valley.

\subsection{Large Field Inflation}
\label{subsec:particlelarge}

One can generate large-scale magnetic field in large field inflation
models if gauge kinetic function has appropriate dependence on the
inflaton field given by Eq.~(\ref{eq:couplinglargefield}). As the model
of inflation we adopt a chaotic inflation in supergravity based on a
shift symmetry $\Phi \longrightarrow \Phi+iC$ where $C$ is a real
number.  Specifically we adopt a model proposed in
Ref.~\cite{Kawasaki:2000yn} with two chiral superfields $\Phi$ and $X$.
In this model by virtue of this symmetry and its soft breaking, the
K\"ahler potential and superpotential are given by
\begin{equation}
K=\frac{1}{2}(\Phi+\Phi^\ast)^2+XX^\ast+\cdots\, ,\quad W=mX\Phi,
\end{equation}
where
$m$ is a constant corresponding to the inflaton mass. The scalar
Lagrangian possesses standard kinetic terms and is given by 
\begin{equation}
{\mathcal{L}}_{\mathrm{scalar}}=-\partial_\mu\Phi\partial^\mu\Phi^\ast
-\partial_\mu X\partial^\mu X^\ast-V(\Phi,X)\, ,
\end{equation}
where the potential reads
\begin{eqnarray}
V(\Phi,X)&=& m^2{\rm e}^K\Biggl\{
|\Phi|^2\left(1+|X|^4\right)\nonumber \\& &+|X|^2\left[1-|\Phi|^2+
\left(\Phi+\Phi^\ast\right)^2
\left(1+|\Phi|^2\right)\right]\Biggr\}\, . 
\end{eqnarray}
Decomposing the complex scalar field $\Phi$ into two real scalar fields
as $\Phi=(\sigma+i\phi)/\sqrt{2}$ and using the fact that only $\phi$
can have a large initial value beyond the Planck scale without any
exponential barrier due to the K\"ahler potential, we find that the
Universe evolves as in the chaotic inflation model with a massive scalar
field~\cite{Linde:1984st}, namely
\begin{equation}
{\mathcal{L}}_{\mathrm{inflaton}}=-\frac{1}{2}\partial_{\mu}
\phi\partial ^{\mu}\phi-V(\phi)\, ,\quad 
V(\phi)=\frac{1}{2}m^2\phi^2\, .  
\end{equation}
Therefore, as announced, one has obtained the
potential~(\ref{eq:largepot}) with $p=2$. Here, we recall that all
models such that $p>3.1$ are now excluded at $95\%$ confidence level by
the WMAP3 data, see Ref.~\cite{Martin:2006rs}.

\par

We now consider the vector part of the Lagrangian. Specifically we take
the following Lagrangian for the U(1) gauge field 
\begin{equation}
{\mathcal{L}}_{_{\mathrm{EM}}} =-\frac{1}{4}\Re
\tilde{f}(\Phi)F_{\mu\nu}F^{\mu\nu}\, ,
\end{equation}
with
\begin{equation}
\label{eq:choiceflarge}
\tilde{f}(\Phi)\equiv
f_0{\rm e}^{\alpha\Phi^2}\, , 
\end{equation}
where $\alpha$ is a constant.  Note that the
shift symmetry of the system is preserved if $F_{\mu\nu}=0$.  As argued
above, since $\sigma$ rapidly relaxes to the origin at the onset of
inflation, the gauge kinetic function then reduces to 
\begin{equation}
\Re \tilde{f}(\Phi)=f_0 {\rm e}^{-\alpha\phi^2/2} \equiv f^2(\phi)\,
,\end{equation} and it reproduces the coupling function given by
Eq.~(\ref{eq:couplinglargefield}) when $p=2$. However, and contrary to
the case of power-law inflation, the choice~(\ref{eq:choiceflarge}) can
no longer be justified from string theory, at least for the simple
favored case mentioned before.

\subsection{Small Field Inflation}
\label{subsec:particlesmall}

Generation of large-scale magnetic field is also possible in small field
inflation models.  Here we adopt a model proposed in
Ref.~\cite{Yamaguchi:2000vm} which does not require fine tuning of the
initial conditions. This model is based on the following K\"ahler
potential and the superpotential
\begin{equation}
K = \frac{1}{2}(\Phi + \Phi^{\ast})^{2} + XX^{\ast}\, , 
\quad W=v^{2}X\left(1-g\Phi^{2}\right)\, . 
\end{equation}
Symmetry argument to obtain the above expression has been fully
described in Ref.~\cite{Yamaguchi:2000vm} and we do not repeat it here.
In this model the real part of the scalar component $\Phi\equiv (\phi +
i \chi)/\sqrt{2}$ plays the role of the inflaton for the small field
inflation while its imaginary part serves as that for chaotic inflation
which occurs beforehand.  The quantity $X$ is another chiral superfield
introduced to yield an appropriate scalar potential. The scalar part of
the Lagrangian then reads 
\begin{equation}
\mathcal{L}(\phi,\chi,X) =
\frac{1}{2}\partial_{\mu}\phi\partial^{\mu}\phi +
\frac{1}{2}\partial_{\mu}\chi\partial^{\mu}\chi +
\partial_{\mu}X\partial^{\mu}X^{*} -V(\phi,\chi,X)\, , 
\end{equation}
with the potential $V(\phi,\chi,X)$ given by the following expression
\begin{eqnarray}
V(\phi,\chi,X) &=& v^{4} {\rm e}^{|X|^{2}+\phi^{2}}
\Biggl\{\left[\left(1-\frac{g}{2}\phi^{2} \right)^{2} 
+g\chi^{2} 
\left(1+\frac{g}{2}\phi^{2}+\frac{g}{4}\chi^{2}
\right)\right]
\nonumber \\ & & \times
\left(1-|X|^{2}+|X|^{4}\right) 
+|X|^{2} \Biggl[2g^{2}\chi^{2}+2(g-1)^{2}\phi^{2}
\nonumber \\ & &
+2g\left(g+1\right)\phi^{2}\chi^{2}+2g\left(g-1\right)\phi^{4}
+\frac{g^{2}}{2}\phi^{2}\left(\phi^{2}+\chi^{2}\right)^{2}
\Biggr]\Biggr\} \, .
\end{eqnarray}
Since $\chi$ is free from exponential rise as in the model discussed in
the previous subsection it can naturally induce chaotic
inflation. Meanwhile $\phi$ and $X$ settle to the origin apart from
quantum fluctuations. After chaotic inflation $\phi$ is still close to
the origin and then induces new inflation. In this regime with
$\chi=|X|= 0$ and $|\phi|\ll 1$, the scalar potential reads
\begin{equation}
V(\phi)= v^{4}\left(1- \frac{c}{2}\phi^{2}\right), 
\label{eq:potnew}
\end{equation}
with $c \equiv 2(g-1)$. Thus, if $g \gtrsim 1$, $\phi$ rolls down slowly
toward the vacuum expectation value $\phi_0 \equiv \sqrt{2/g}$ and small
field inflation takes place. The potential (\ref{eq:potnew}) has the
same form as (\ref{eq:potsmallhybrid}).

\par

Finally, using Eq.~(\ref{eq:couplingnew}), the desired form of the gauge
kinetic function is realized if we take 
\begin{equation}
\tilde{f}(\Phi)\equiv f_0\Phi^{2\alpha/c}\, .
\end{equation}
The branch $\alpha \sim 2$ is the one to be used in order to obtain
positive $q$. The favored string inspired model $q=1$ is obtained for
$c\simeq 4$ or $g\simeq 3$. However, unfortunately, this value is not
compatible with the CMB constraints. Indeed, in the case of small field
models, the first slow-roll parameter $\epsilon _1$ is exponentially
small while the second one is given by $\epsilon _2=2c$, see
Ref.~\cite{Martin:2006rs}. The CMB constraints on $\epsilon _2 $ are
$-0.07<\epsilon _2<0.07$ at $95\%$ CL, see Ref.~\cite{Martin:2006rs},
which implies $-0.035<c<0.035 $. This means that $q\gtrsim 57$, a value
not compatible with the string inspired value and too large to be
considered as natural. However, at the same time, the model makes use of
the branch where the back-reaction problem is not severe at all. In this
case, the characteristics of the required reheating stage seems easy to
satisfy.

\section{Conclusion}
\label{sec:conclusion}

In this section, we recap our main results and discuss issues that
should be studied in future works. In the present paper, we have studied
the generation of large-scale magnetic fields in inflationary cosmology,
breaking the conformal invariance of the electromagnetic field by
introducing a coupling with the inflaton field as it is generic in a
supergravity framework. We have determined the form of the coupling
necessary in order to produce large-scale magnetic fields with the
required strengths on the relevant astrophysical scales. We have shown
that the scale-dependent magnetic energy density possess two
branches. Among these two branches, one ($\alpha \sim -3$) seems to lead
to sensible model building in the case of power-law inflation but only
at the expense of having a low energy inflation scale and a long
reheating stage. The other branch ($\alpha \sim 2$) is useful in the
case of small field models and this case appears to be relevant since
this does not require to fine-tune the reheating epoch. However, the
model building condition $q\gtrsim 57$ cannot be justified in a string
inspired model and seems artificial. Finally, in the context of large
field models, we have shown that the required coupling is quite
difficult to justify.

\par

Determining exactly all the consistent models in the framework envisaged
here is a non-trivial issue because the parameter space is large (at
least four-dimensional), in particular when a general reheating stage is
considered. Here, we have just proven that consistent models
exist. However, it would certainly be interesting to systematically
explore the parameter space in order to have a more accurate idea of
whether the consistent models are just peculiar or, on the contrary,
quite generic. Another interesting avenue for the future is clearly the
model building issue. Here, we have noticed that the branch $\alpha \sim
2$ can be used only in the context of small field models. It would be
interesting to find other models where this can also been done. We hope
that these issues will be addressed in the near future.

\acknowledgments

This work was partially supported by CNRS-JSPS Bilateral Joint Project
``The Early Universe: a precision laboratory for high energy
physics.''  The work of J.Y.  was partially supported by the JSPS
Grant-in-Aid for Scientific Research Nos.~16340076 and 19340054.

\section*{References}

\bibliography{references}

\end{document}